\documentclass[11pt,numbers,sort&compress]{article}

\usepackage{latexsym,amsmath,amssymb,theorem,epsfig} 
\usepackage{natbib}
\usepackage{graphicx}

\DeclareFontFamily{OT1}{rsfs10}{}
\DeclareFontShape{OT1}{rsfs10}{m}{n}{ <-> rsfs10 }{}
\DeclareMathAlphabet{\mathscript}{OT1}{rsfs10}{m}{n}

\newcommand{\be}{\begin{equation}}
\newcommand{\ee}{\end{equation}}
\newcommand{\nn}{\nonumber}
\newcommand{\bea}{\begin{eqnarray}}
\newcommand{\eea}{\end{eqnarray}}
\newcommand{\ba}{\begin{array}}
\newcommand{\ea}{\end{array}}

\newcommand{\tr}{\textrm{tr}}

\newcommand{\pt}{\partial}

\def\cX{{\cal X}}
\def\cZ{{\cal Z}}
\def\a{\alpha}
\def\b{\beta}

\def\d{\delta}

\def\k{\kappa}

\def\w{\wedge}

\def\cG{{\cal G}}

\def\cK{{\cal K}}

\begin{document}

\begin{titlepage}
  \vspace*{\stretch{1}}
  \begin{center}
     \Large Flux, Gaugino Condensation and Anti-Branes in Heterotic M-theory
  \end{center}
  \vspace*{\stretch{2}}
  \begin{center}
    \begin{minipage}{\textwidth}
      \begin{center}
        James Gray${}^1$,
        Andr\'e Lukas${}^2$ and
        Burt Ovrut${}^3$
      \end{center}
    \end{minipage}
  \end{center}
  \vspace*{1mm}
  \begin{center}
    \begin{minipage}{\textwidth}
      \begin{center}
        ${}^1$Institut d'Astrophysique de Paris and APC, 
        Universit\'e de Paris 7,\\
        98 bis, Bd.~Arago 75014, Paris, France\\[0.2cm]
        ${}^2$Rudolf Peierls Centre for Theoretical Physics, 
        University of Oxford,\\ 
        1 Keble Road, Oxford OX1 3NP, UK\\[0.2cm]
        ${}^3$Department of Physics, University of Pennsylvania,\\
        Philadelphia, PA 19104--6395, USA
      \end{center}
    \end{minipage}
  \end{center}
  \vspace*{\stretch{1}}
\begin{abstract}
  \normalsize We present the potential energy due to flux and gaugino
  condensation in heterotic M-theory compactifications with
  anti-branes in the vacuum. For reasons which we explain in detail,
  the contributions to the potential due to flux are not modified from those
  in supersymmetric contexts. The discussion of gaugino condensation
  is, however, changed by the presence of anti-branes. We show how a
  careful microscopic analysis of the system allows us to use standard
  results in supersymmetric gauge theory in describing such effects -
  despite the explicit supersymmetry breaking which is
  present. Not surprisingly, the significant effect of anti-branes on the
  threshold corrections to the gauge kinetic functions greatly alters
  the potential energy terms arising from gaugino condensation.
\end{abstract}
  \vspace*{\stretch{5}}
  \begin{minipage}{\textwidth}
    \underline{\hspace{5cm}}
    \\
    \footnotesize
    ${}^1$email: gray@iap.fr \\
    ${}^2$email: lukas@physics.ox.ac.uk \\
    ${}^3$email: ovrut@elcapitan.hep.upenn.edu
  \end{minipage}
\end{titlepage}


\section{Introduction}

In \cite{paper1}, the perturbative four-dimensional effective action
of heterotic M-theory compactified on vacua containing both branes and
anti-branes in the bulk space was derived. That paper concentrated
specifically on those aspects of the perturbative low energy theory
which are induced by the inclusion of anti-branes. Hence, for clarity
of presentation, the effect of background $G$-flux on the effective
theory was not discussed. Furthermore, non-perturbative physics,
namely gaugino condensation and membrane instantons, was not included.
However, all three of these contributions are required for a complete
discussion of moduli stabilization, ${\cal{N}}=1$ supersymmetry
breaking and the cosmological constant. Therefore, in this paper, we
extend the results of \cite{paper1} to include the effects of flux and
gaugino condensation.  Another piece of the effective theory, that is,
the contribution of membrane instantons, will be presented elsewhere
\cite{paper3}. There is of course a vast literature on the subject of
moduli stabilization, flux and gaugino condensation in heterotic
theories. Some recent discussions of various aspects of these topics
appear in
\cite{Buchbinder:2003pi,Gukov:2003cy,Becker:2004gw,Gurrieri:2004dt,Curio:2005ew,deCarlos:2005kh,Anguelova:2006qf,Ovrut:20061,Serone:2007sv,Correia:2007sv}.

A strong motivation for attempting to find stable vacua in Calabi-Yau
compactifications of heterotic theories comes from the advantages that
such constructions enjoy in particle physics model building.  For
example, models with an underlying ${\rm SO}(10)$ GUT symmetry can be
constructed where one right handed neutrino per family occurs
naturally in the $\bf 16 \rm$ multiplet and gauge unification is
generic due to the universal gauge kinetic functions in heterotic
theories. Recent progress in the understanding of non-standard
embedding models~\cite{A1,A2,A3,A4} and the associated mathematics of
vector bundles on Calabi-Yau spaces~\cite{B1,B2,B3,B4} has led to the
construction of effective theories close to the Minimal Supersymmetric
Standard Model (MSSM), see~\cite{Donagi:2004ub, Braun:2005ux,
  Braun:2005bw, Braun:2005zv, Braun:2006ae1, Braun:2006ae}.  This has
opened up new avenues for heterotic phenomenology. For example, one
can proceed to look at more detailed properties of these models such
as $\mu$ terms~\cite{Braun:2006ae2}, Yukawa
couplings~\cite{Braun:2006ae3}, the number of
moduli~\cite{Braun:2006ae4} and so forth. Other groups are also
making strides in heterotic model building, see for example
\cite{Bouchard:2005ag,Bouchard:2006dn,Andreas:2006dm,Faraggi:2006qa,Lebedev:2006kn,Blumenhagen:2006wj}.

The addition of $G$-flux to the formalism described in \cite{paper1}
is, as we will show in this paper, relatively straightforward. In
contrast, it is not at first obvious how to incorporate gaugino
condensation into this explicitly non-supersymmetric compactification
of M-theory. The reason is that almost everything we know about this
non-perturbative phenomenon is based on the dynamics of unbroken
${\cal N}=1$ supersymmetric gauge theories. However, by carefully
analyzing the limit where the usual discussions of gaugino
condensation are applied, we will show that our non-supersymmetric
system reverts to a globally supersymmetric gauge theory. This fact
allows us to construct the condensation induced potential energy terms
in the presence of anti-branes using a component action approach
similar to that of \cite{Dine:1985rz}.

The plan of this paper is as follows. In the next section, we briefly
review those aspects of \cite{paper1} which are required for the
current work. In Section \ref{susyflux}, we review the subject
of flux in supersymmetric heterotic M-theory. In Section
\ref{fluxsection}, the effects of $G$-flux in heterotic M-theory vacua
which include anti-branes are explicitly computed. Gaugino
condensation in the presence of anti-branes is then introduced in
Section \ref{sectiongaugino}. We begin by describing how the potential energy
terms induced by this effect can be calculated in terms of the
condensate itself. It is then shown how one can explicitly evaluate
the condensate as a function of the moduli fields.  Finally, we
conclude in Section \ref{conclusion} by writing out, in full, the
effective potential energy we have obtained by compactifying heterotic
M-theory in the presence of anti-branes, $G$-flux and gaugino
condensation. 
\section{Anti-Branes in Heterotic M-Theory}
\label{ab}

\begin{figure}[ht]\centering 
\includegraphics[height=9cm,width=14cm, angle=0]{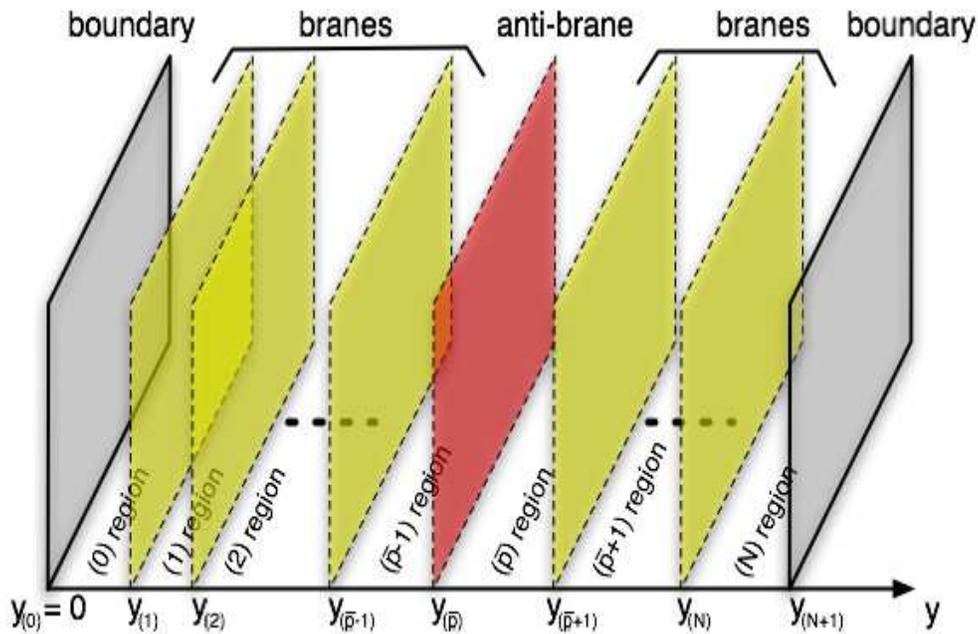}
\caption{The brane configuration in five-dimensional heterotic
  M-theory}
\label{fig1}
\end{figure}

The low energy effective Lagrangian of heterotic M-theory 
compactified on vacua containing bulk space branes and anti-branes, 
but ignoring $G$-flux and non-perturbative effects, was constructed in 
\cite{paper1}.  In this section, we provide a brief
summary of the aspects of \cite{paper1} which are important in the
current paper. The basic vacuum configuration which we consider, 
as viewed from five dimensions, is depicted in Figure \ref{fig1}.
We include an arbitrary number of three-branes in the vacuum, but, for
clarity of notation, restrict the discussion to a single anti
three-brane. Our results extend almost trivially to the case where
multiple anti-branes are present.  The (anti) three-branes arise as
the low energy limit of (anti) M5-branes wrapped on holomorphic curves
in the internal Calabi-Yau threefold. We will often refer to the
(anti) three-branes simply as (anti-) branes.  A more detailed
discussion of this setup is provided in \cite{paper1}.  As shown
in Figure \ref{fig1}, we label the extended objects with a bracketed
index $(p)$ ranging from $(0)$ to $(N+1)$. The values $(0)$ and
$(N+1)$ correspond to the orbifold fixed planes, $(\bar{p})$ labels
the anti-brane and the remaining $(p)$ values are associated with the
branes. The bulk regions are labeled by the same index as the brane
which borders them on the left. Fields associated with the world
volume of a given extended object or with a certain region of the bulk
are often labelled with the associated index.

In \cite{paper1}, our starting point was the five-dimensional action
describing the compactification of Ho\v{r}ava-Witten theory on a
Calabi-Yau threefold $X$ in the presence of both M5 branes and an anti-M5
brane.  The bosonic field content of this theory is as follows. Let
$\alpha,\beta, \ldots =0,\dots,4$ be the five-dimensional bulk space
indices, $\mu,\nu,\ldots=0,\dots,3$ label the four-dimensional
Minkowski space indices and $y$ be the coordinate of the
$S^{1}/{\mathbb Z_{2}}$ orbifold interval.  Additionally, we let 
$i,j,k,\ldots=1,\dots,h^{1,1}$, $\underline{a},\underline{b},\ldots=1,\dots,h^{2,1}$ and 
$\underline{A},\underline{B},\ldots=1,\dots,h^{2,1}+1$, where $h^{1,1}$ and $h^{2,1}$ are the 
dimensions of the $H^{1,1}(X)$ and $H^{2,1}(X)$ cohomology groups respectively.
Then, in the bulk space we
have the graviton, $g_{\alpha \beta}$, $h^{1,1}$ Abelian vector fields
${{\cal A}_{\alpha}}^k$ with field strengths ${{\cal
    F}_{\alpha\beta}}^k$, a real scalar field $V$, $h^{1,1}$ real
scalar fields $b^k$ which obey the condition $d_{i j k} b^i b^j b^k
=6$ (the $d_{ijk}$ being the intersection numbers on the Calabi-Yau threefold) and,
therefore, constitute $h^{1,1}-1$ degrees of freedom, $h^{2,1}$
complex scalar fields ${\mathfrak z}^{\underline{a}}$, $2(h^{2,1}+1)$ real scalar
fields ${\xi}^{\underline{A}}$, $\tilde{\xi}_{\underline{B}}$ with their field strengths
${{\cal{X}}^{\underline{A}}}_{\alpha}$, ${{\cal{X}}_{\underline{B}}}_{\beta}$ and the
three-form $C_{\alpha \beta \gamma}$ with its field strength
$G_{\alpha \beta \gamma \delta}$. Of these bulk fields, $V$,
${\mathfrak z}^{\underline{a}}$, $g_{\mu \nu}$, $g_{yy}$, $b^k$, $C_{\mu \nu y}$ and
${\cal A}^k_y$ are even under the ${\mathbb{Z}}_2$ orbifold
projection, while $g_{\mu y}$, $\xi^{\underline{A}}$, $\tilde{\xi}_{\underline{B}}$, ${\cal
  A}^k_{\mu}$, $C_{\mu \nu \gamma}$ are odd.

In addition, there are extra degrees of freedom living on the extended
sources in the vacuum. On each of the two four-dimensional fixed
planes, one finds ${\cal N}=1$ gauge supermultiplets.  These contain
gauge fields $A_{(p)\mu}$ indexed over the adjoint representation of
the unbroken gauge group ${\cal{H}}_{(p)} \subseteq E_{8}$ for
$p=0,N+1$.  The associated fields strengths are denote by
$F_{(p)\mu\nu}$. Furthermore, on the fixed planes there are ${\cal
  N}=1$ chiral matter supermultiplets with scalar components
$C_{(p)}^{Rx}$, $p=0,N+1$ transforming in various representations $R$,
with components $x$, of this gauge group. Details of the origin and
structure of the matter sector can be found in the Appendix of
\cite{paper1}.  The world volume fields associated with the
three-branes are as follows. First, each brane $(p)$ has an embedding
coordinate, that is, the brane position, $y_{(p)}$ and a world volume
scalar $s_{(p)}$.  Second, each brane supports an ${\cal N}=1$ gauge
supermultiplet with structure group $U(1)^{g_{(p)}}$ where $g_{(p)}$
is the genus of the curve wrapped by brane $(p)$. These contain the
Abelian gauge fields $A_{(p)\mu}^{u}$, where $u=1,\dots,g_{(p)}$. The
associated field strengths are denoted by $E^{u}_{(p)}$. In general,
there will be additional chiral multiplets describing the moduli space
of the curves around which the M5 and anti-M5 branes are
wrapped. Furthermore, there may be non-Abelian generalizations of the
gauge field degrees of freedom when branes are stacked. However, these
multiplets are not vital to our discussion and, therefore, we shall
not explicitly take them into account.

\vspace{0.1cm}

Given this field content, the five-dimensional action
describing Ho\v{r}ava-Witten theory compactified on an arbitrary Calabi-Yau
threefold in the presence of M5 branes and an anti-M5 brane is given by

\bea \label{5daction}\nonumber S = - \frac{1}{2 \kappa_5^2} \int d^5 x
\sqrt{-g} \left[ \frac{1}{2} R + \frac{1}{4} G_{kl}(b) \partial
  b^k \partial b^l + \frac{1}{2} G_{kl}(b) {\cal F}^k_{\alpha \beta}
  {\cal F}^{l \alpha \beta} + \frac{1}{4} V^{-2} (\partial V)^2 +
  \lambda (d_{i j k} b^i b^j b^k -6)\right. \\ \nonumber \left. +
  \frac{1}{4} {\cal K}_{\underline{a} \underline{\bar{b}}}({\mathfrak z })\partial {\mathfrak
    z}^{\underline{a}}
  \partial \bar{{\mathfrak z}}^{\underline{\bar{b}}} - V^{-1} (\tilde{\cal
    X}_{\underline{A}\alpha} - \bar{M}_{\underline{A}\underline{B}} ({\mathfrak z}){\cal X}_{\alpha}^{\underline{B}}) (
  [ \textnormal{Im}{(M ({\mathfrak z}))}]^{-1})^{\underline{A}\underline{C}} (\tilde{\cal
    X}_{\underline{C}}^\alpha - M_{\underline{C}\underline{D}} ({\mathfrak z}){\cal X}^{\underline{D}\alpha}) \right.\\
\nonumber \left. +\frac{1}{4!} V^2 G_{\alpha \beta \gamma \delta}
  G^{\alpha \beta \gamma \delta} + m^2V^{-2} G^{kl} (b) \hat{\beta}_k
  \hat{\beta}_l \right] \\ \nonumber -\frac{1}{2 \kappa_5^2} \int
\left( \frac{2}{3} d_{klm} {\cal A}^k \wedge {\cal F}^l \wedge {\cal
    F}^m + 2 G \wedge (( \xi^{\underline{A}} \tilde{\cal X}_{\underline{A}} - \tilde{\xi}_{\underline{A}}{\cal
    X}^{\underline{A}}) - 2 m\hat{\beta}_k {\cal A}^k) \right)\eea

\bea \label{action} - \int d^5 x \; \delta (y)
 \sqrt{- h_{(0)} }\,\left[ \frac{m}{\kappa_5^2} V^{-1} b^k \tau^{(0)}_k +
  \frac{1}{8 \pi \alpha_{\textnormal{GUT}}} V \rm{tr}(F_{(0)}^2) +
  {\cal G}_{(0)RS} D_{\mu} C^{Rx}_{(0)} D^{\mu} \bar{C}^S_{(0) x}
\right. \\ \nonumber \left. + V^{-1} {\cal G}^{RS}_{(0)} \frac{\partial
    W_{(0)}}{\partial C^{Rx}_{(0)}} \frac{\partial
    \bar{W}_{(0)}}{\partial \bar{C}^S_{(0) x}} + {\rm tr}(D_{(0)}^2) \right] \eea

\bea \nonumber -  \int d^5 x \; \delta (y - \pi
\rho) \sqrt{- h_{(N+1)} }\,\left[\frac{m}{\kappa_5^2} V^{-1} b^k \tau^{(N+1)}_k +
  \frac{1}{8 \pi \alpha_{\textnormal{GUT}}} V {\rm tr}(F_{(N+1)}^2)
  + {\cal G}_{(0)RS}
  D_{\mu} C^{Rx}_{(N+1)} D^{\mu} \bar{C}^S_{(N+1) x} \right. \\
\nonumber \left.+ V^{-1} {\cal G}^{RS}_{(0)} \frac{\partial W_{(N+1)}}{\partial
    C^{Rx}_{(N+1)}} \frac{\partial \bar{W}_{(N+1)}}{\partial
    \bar{C}^S_{(N+1) x}} + {\rm tr}(D_{(N+1)}^2) \right]
\eea

\bea \nonumber - \frac{1}{2 \kappa_5^2} \int d^5 x
\;\sum_{p=1}^{N} (\delta (y- y_{(p)})+\delta (y+y_{(p)}))
\left\{\sqrt{-h_{(p)}}\,\left[ mV^{-1} \tau^{(p)}_k b^k + \frac{2 (
      n_{(p)}^k \tau^{(p)}_k)^2}{V (\tau^{(p)}_l b^l)} j_{{(p)}
      \mu} j^{\mu}_{(p)} \right. \right. \\ \nonumber \left. \left. +
    [ \textnormal{Im}{\Pi}]_{{(p)} uw} E^u_{{(p)} \mu \nu} E^{w \mu
      \nu}_{(p)} \right] - 4 \hat{C}_{(p)} \wedge \tau^{(p)}_k
  d(n^k_{(p)} s_{(p)}) - 2 [ \textnormal{Re}{ \Pi}]_{(p)uw} E^u_{(p)}
  \wedge E^w_{(p)} \right\}. \eea
Here $\kappa_{5}$ and $\alpha_{GUT}$ are given by
\begin{equation}
\kappa_{5}^{2}=\frac{\kappa_{11}^{2}}{v}, \quad \alpha_{GUT}
=\frac{(4\pi \kappa_{11}^{2})^{2/3}}{v},
\label{A}
\end{equation}
where $\kappa_{5}$ and $\kappa_{11}$ are $5$- and $11$-dimensional 
Planck constants respectively, $v$ is the reference Calabi-Yau volume and
constant $m$ is the reference mass scale
\begin{equation}
\label{BB}
 m=\frac{2 \pi}{v^{\frac{2}{3}}}
 \left( \frac{\kappa_{11}}{4 \pi} \right)^{\frac{2}{3}}.
\end{equation}
The above expressions describe, in the order presented, the bulk space, the
fixed planes and the world volumes of the M5 and anti-M5 branes. 
The metrics $G_{kl}(b)$, ${\cal{K}}_{\underline{a}\underline{\bar{b}}}({\mathfrak z})$
as well as the matrix $M_{\underline{A}\underline{B}}({\mathfrak z})$  appearing in the bulk space term are defined 
in Appendix A of \cite{paper1}. Their precise form is not required in this paper. The $d_{klm}$
coefficients are the intersection numbers of the Calabi-Yau threefold. The
integers $\tau_k^{(p)}$ are the tensions of the branes, anti-branes
and fixed planes. Charges on individual branes are simply labeled by
$\beta^{(p)}_{k}$ and are equal to the associated tensions in the case of
branes and fixed planes and equal in magnitude, but opposite in sign, in
the case of the anti-brane. The anti-brane is taken to be associated with a purely
anti-effective curve for simplicity. The $\hat{\beta}_k$ coefficients are the
sum of the charges on all extended sources to the left of where
the bulk theory is being considered. We will frequently denote
quantities associated with the anti-brane by a bar. Hence, the tension
of the anti-brane is denoted by $\bar{\tau}_k =\tau^{(\bar{p})}_{k}$.
The metrics ${\cal{G}}_{(p)RS}$, the matter field superpotentials $W_{(p)}$ and the D-terms $D_{(p)}$ with $p=0,N+1$ appearing in the boundary plane terms are defined in Appendix B of \cite{paper1}. Again, their precise form is not required for the analysis in this paper. 
The quantities
\begin{equation}
n_{(p)}^{k}=\frac{\beta_{k}^{(p)}}{\Sigma_{l=1}^{h^{1,1}}\beta_{l}^{(p)2}}
\label{today}
\end{equation}
are the normalized version of the three-brane charges and
 we have used the definition  
\bea j_{(p) \mu} &=&
\frac{\beta^{(p)}_{k}}{n_{(p)}^l \beta^{(p)}_{l}} ( d(n^k_{(p)} s_{(p)}) -
\hat{{\cal A}}^k_{(p)} )_{\mu}. \eea Hats on field quantities denote
pullbacks of the corresponding bulk variable onto the extended object
under consideration. The matrix $\Pi_{(p)uw}$ appearing in the world volume brane terms
is the period matrix of the curve on which the $p$-th brane or anti-brane is wrapped. This is defined
in Appendix C of \cite{paper1}.
Finally, the quantities $h_{(p)}$ are the induced metrics on the 
boundary planes, branes and the anti-brane.
It is important to note that despite the appearance of an anti M5-brane in the vacuum, action~\eqref{5daction} is ${\cal{N}}=1$ supersymmetric, both in the five-dimensional bulk space and on all four-dimensional fixed planes and branes. The reason for this is that the dimensional reduction was carried out with respect to the supersymmetry preserving Calabi-Yau threefold, and did not explicitly involve the anti M5-brane. However, as we now briefly discuss, the anti M5-brane {\it does} enter the dimensional reduction of the theory to four-dimensions, explicitly breaking its ${\cal{N}}=1$ supersymmetry.

\vspace{0.1cm}

In \cite{paper1}, we described how the presence of an anti-brane in
the bulk space of a heterotic M-theory vacuum changes the warping in
the extra dimensions. We showed that this back-reaction is manifested
by a quadratic contribution to the warping.  It is interesting to note
that this contribution derives simply from the failure of the tensions
of the extended objects to sum to zero in the case where anti-branes
are present. This property of the source terms is responsible for all
of the major changes to the perturbative theory. The only bulk fields
involved in the generalized domain wall solution including an
anti-brane are the metric $g_{\alpha\beta}$, the volume modulus $V$
and the K\"ahler moduli $b^{k}$. We take as our metric ansatz
\be \label{metric} ds^2_5 = a(x^{\mu},y)^2 g_{4\mu \nu}(x^{\mu}) d
x^{\mu} d x^{\nu} + b(x^{\mu},y)^2 d y^2. \ee It turns out to be
helpful to introduce a function which encodes the standard linear
warping functions of heterotic M-theory in a usefully normalized
manner. It is given by
\begin{equation}
h_{(p)k}(z) = \sum^{p}_{q=0} \tau^{(q)}_k (z-z_{(q)}) - \frac{1}{2}
\sum_{q=0}^{N+1} \tau^{(q)}_k z_{(q)}(z_{(q)} - 2) - \delta_k\; .
\label{B}
\end{equation}
The $z$ which appears here is defined by $z = y/ \pi \rho$, where $\pi
\rho$ is the reference length of the orbifold interval and 
$y$ is the coordinate of the fifth dimension. Each $z_{(q)}$  is the $\pi\rho$ normalized 
position modulus for the $q$-th brane or anti-brane.  Note that the fixed planes are located at 
$z_{(0)}=0$ and $z_{(N+1)}=1$ respectively. In expression \eqref{B}, 
\begin{equation}
\delta_k = \frac{1}{2} (\bar{\tau}_k -
\bar{\beta}_k)=\bar{\tau}_k.
\label{CCC}
\end{equation}
The quantity $h_{(p)k}$ has been defined so that its orbifold average is zero. This property
enables us to extract definitions of the zero modes of the
compactification which  lead to a four-dimensional action in a
sensible form, without the need to use field redefinitions.

Using these definitions and the equations of motion for $V$
and $b^k$, we find the following non-supersymmetric domain wall
ansatz around which the theory will be reduced to four-dimensions.
It is given by
\begin{eqnarray}
  \frac{a_{(p)}}{a_0}&=&1-\epsilon_0\frac{b_0}{3V_0}b_0^k\left[ h_{(p)k}+
    \delta_k\left( z^2-\frac{1}{3}\right)\right]\label{a}\\
  \frac{V_{(p)}}{V_0}&=&1-2\epsilon_0\frac{b_0}{V_0}b_0^k\left[ h_{(p)k}
    -\delta_k
    \left( z^2-\frac{1}{3}\right)\right]\label{V}\\
  b_{(p)}^k&=&b_0^k+2\epsilon_0
  \frac{b_0}{V_0}\left[\left(h_{(p)}^k-\frac{1}{3}h_{(p)l}b_0^kb_0^l
    \right)-\left( \delta^k-\frac{1}{3}\delta_lb_0^kb_0^l
    \right)\left( z^2-\frac{1}{3}\right)\right],\label{bk}
\end{eqnarray}
where 
\begin{equation}
\label{AA}
\epsilon_{0}=\pi\rho m
\end{equation}
with $m$ defined in \eqref{BB}.  The remaining notation is as
explained earlier, with the one addition that the ``$0$'' subscript on
$a_{0}$, $b_{0}$, $V_{0}$ and $b^{k}_{0}$ denotes quantities which are
functions of the four uncompactified coordinates only. It is these
functions that become the moduli when we dimensionally reduce on this
ansatz to obtain the effective four-dimensional action. Note that we
have not given an expression for the warping of the metric coefficient
$b$. This is because this $y$ dependence can be removed by a
coordinate redefinition and, hence, does not enter the calculation of
the four-dimensional effective action.

We emphasize several points about this result which will be important
in this paper. First, the $z$ independent factors have been
defined such that $a_0$, $b_{0}$, $V_0$ and $b^k_0$ are simply the 
orbifold averages of $a$, $b$, $V$ and $b^k$ respectively. This proves to be useful in
performing the dimensional reduction around this
configuration. Second, upon taking $\delta_k \rightarrow 0$ one
recovers the conventional results for heterotic M-theory, given, for
example, in \cite{mattias}. In particular, the quadratic pieces of the
warping correctly disappear as we turn the anti-brane into an M5 brane
in this manner. Note that in each of the above expressions the warping
occurs with the same multiplicative factor, specifically
$\epsilon_{0}\frac{b_0}{V_0}$. In \cite{paper1,Lukas:1998hk}, it was shown that 
the four-dimensional effective theory is an expansion in the parameters 
$\epsilon_S$ and $\epsilon_R$ given by
\be \epsilon_S = \pi \left( \frac{\kappa_{11}}{4 \pi}
\right)^{\frac{2}{3}} \frac{2 \pi \rho}{v^{\frac{2}{3}}}
\frac{b_0}{V_0} \;\;\;\; , \;\;\;\; \epsilon_R =
\frac{v^{\frac{1}{6}}}{\pi \rho} \frac{V_0^{\frac{1}{2}}}{b_0}.
\label{x}
\ee 
The first of these is the ``strong coupling'' parameter which specifies the size of the warping in the 
orbifold dimension. The second  term is the expansion parameter which controls
the size of the Calabi-Yau Kaluza-Klein modes. Using~\eqref{BB} and~\eqref{AA}, we see that 
\begin{equation}
\epsilon_{S}=\epsilon_{0}\frac{b_0}{V_0},
\label{good}
\end{equation}
as it should be.

Further results from \cite{paper1} will be reproduced as and when
required in the following sections. 

\section{Flux in Supersymmetric Heterotic M-Theory}
\label{susyflux}

In this section, we review the subject of flux in supersymmetric
Calabi-Yau compactifications of heterotic M-theory. Initially, we will
consider heterotic theory from an eleven-dimensional perspective so
that a complete understanding is obtained of all of the different
fluxes present, including those whose nature is somewhat obscured by
the five-dimensional description. Later on in this section, however,
having gained this understanding, we revert to the five-dimensional
picture which is technically more convenient for the discussion of the
addition of anti-branes. The full eleven-dimensional $M$-theory
indices will be denoted by $I,J,K,\ldots=0,\dots, 9,11$, whereas
ten-dimensional indices orthogonal to the orbifold direction $y$ are
written as $\bar{I},\bar{J},\bar{K},\ldots=0,\dots,9$. Furthermore,
the six-dimensional real coordinates on the Calabi-Yau manifold
$(CY_{3})$ will be specified by $A,B,C,\ldots=4,\dots,9$. The
associated holomorphic and anti-holomorphic complex coordinates on the
threefold are then written as $a,b,c,\ldots=1,2,3$ and
$\bar{a},\bar{b},\bar{c},\ldots=1,2,3$ respectively. Finally, the
indices $\hat{i},\hat{j}, \hat{k},\ldots=4,\dots,9,11$ run over both
the $CY_3$ and orbifold directions. The material on the non-zero mode
of heterotic M-theory reviewed here was first presented in
\cite{Witten:1996mz,Lukas:1997fg,Lukas:1998tt,Lukas:1998hk}.

\subsection{The components of $G$ relevant to stable vacua}
\label{fluxcomp}

Let us find the components of the eleven-dimensional background four-form flux $G_{IJKL}$  that can be non-vanishing in a realistic, stable vacuum. To begin, note that there are three possible types of index structure for the four-form $G$.

\begin{itemize}
\item {\bf One or more 4D indices:} The possibilities are $G_{\mu \nu \gamma
    \delta}, G_{\mu \nu \gamma \hat{i}}, G_{\mu \nu \hat{i} \hat{j}}$
  and $G_{\mu \hat{i} \hat{j} \hat{k}}$.
 \begin{itemize}
 \item $G_{\mu \nu \gamma \hat{i}}, G_{\mu \nu \hat{i} \hat{j}}$ and
   $G_{\mu \hat{i} \hat{j} \hat{k}}$ all obviously break maximal spacetime
   symmetry in four dimensions (that is, Poincar$\rm \acute{e}$, de Sitter or 
   anti-de Sitter symmetry) if present in the vacuum. As such, they are
   not relevant for a discussion of realistic stabilized vacua and will be set to zero.
 \item A priori, $G_{\mu \nu \gamma \delta}$ could have an expectation
   value proportional to the four-dimensional volume element without
   breaking maximal spacetime symmetry. However, $G$ is intrinsically
   odd under the Ho\v{r}ava-Witten $\mathbb{Z}_2$ and so, since
   $G_{\mu \nu \gamma \delta}$ has no $y$ index, this component is
   also odd under the orbifolding. Given this, if non-zero, $G_{\mu
     \nu \gamma \delta}$ must jump at the orbifold fixed
   planes.  Were there to be such a discontinuity,  the exterior derivative 
   $(dG)_{y \mu \nu \gamma \delta}$ would be a delta function at each fixed plane and, hence,
   require charges of the appropriate index for global consistency. On the
   fixed planes we have two options, either an $(F \wedge F)_{\mu \nu \gamma \delta}$ or
   $(R\wedge R)_ {\mu \nu \gamma \delta} $ source. However, giving an 
   expectation value to a four-dimensional vector gauge
   field clearly breaks our requirement of maximal
   spacetime symmetry. Furthermore, it turns out that $(R \wedge R)_{\mu \nu \gamma \delta}$ 
   for a space of maximal spacetime symmetry vanishes. Therefore,  a purely
   external component of the four-form in a realistic, 
  stable vacuum must also be chosen to be zero. We conclude that, if they are to be non-vanishing, 
  all indices of $G$ must lie on the internal seven-orbifold.
   \end{itemize}

 \item {\bf All indices on the $CY_3$:}  Written in terms of the complex coordinates on the $CY_3$, there are three possible index structures for the $G_{ABCD}$ components; 
 $G_{abc\bar{d}}, G_{\bar{a}\bar{b}\bar{c }d}$ and $G_{a\bar{b}c \bar{d}}$. Since $G$ is intrinsically odd and $G_{ABCD}$ has no $y$ index, all of these components are odd under the $\mathbb{Z}_2$ orbifolding. 

 \item {\bf One index on the orbifold and the rest on the $CY_3$:}
   In this case there are four possible complex index structures, $G_{abcy}, G_{\bar{a}\bar{b}\bar{c}y}$
 and $G_{ab\bar{c}y}, G_{\bar{a}\bar{b}cy}$, for the
   $G_{ABC y}$ components. All of these are, however, even under the
   $\mathbb{Z}_2$ due to the $y$ index which they carry.

\end{itemize}

In summary, if we are interested in stable vacua of heterotic
M-theory with maximal spacetime symmetry, we need only consider four-form fluxes with all indices
internal to the compactification seven-orbifold. Of these, the
$G_{ABCD}$ are odd and the $G_{ABCy}$ even under the $\mathbb{Z}_2$
action.

\subsection{The Bianchi identity and its sources}

The components of $G$ are constrained by the requirement of anomaly freedom to satisfy the Bianchi identity of eleven-dimensional heterotic M-theory with $N$ bulk M5 branes. It follows from the above discussion that the non-vanishing components of $G$ are contained in the seven-dimensional restriction of this identity given by
  \bea \label{BI} (dG)_{y ABCD} &=& 4  \pi
\left(\frac{\kappa_{11}}{4 \pi} \right)^{\frac{2}{3}} \left[ \delta
  (y) J^{(0)} + \delta (y - \pi \rho) J^{(N+1)} + \frac{1}{2}
  \sum_{p=1}^{N} J^{(p)} \left( \delta (y - y_{(p)}) + \delta (y +
    y_{(p)}) \right) \right]_{ABCD} \eea The four-form charges localized on the orbifold fixed
planes and bulk M5 branes in this expression are 
 \bea \label{charges} J^{(0)} &=& -\frac{1}{16
  \pi^2} \left( \tr F^{(1)} \wedge F^{(1)} - \frac{1}{2} \tr R \wedge
  R \right)|_{y=0} \;,\\ \nonumber J^{(N+1)} &=& -\frac{1}{16 \pi^2}
\left( \tr F^{(2)}
  \wedge F^{(2)} - \frac{1}{2} \tr R \wedge R \right)|_{y=\pi \rho} \;,\\
\nonumber J^{(p)} &=& \delta ( {\cal C}^{(p)}_2 ) \; \eea respectively, 
where $\delta ({\cal C}^{(p)}_2 )$ is a delta-function four-form localized on the
curve ${\cal C}^{(p)}_2$ wrapped by the $p$-th M5 brane.  For
any two-form $\chi$, we have $\int_{CY_3} \chi \wedge \delta ({\cal
C}^{(p)}_2) = \int_{{\cal C}^{(p)}_2} \chi$. Note that since $d(dG)=0$, it
follows that the sources $J^{(p)}$,$p=0,\dots,N+1$ are closed four-forms on the $CY_{3}$.

Integrability conditions arise from the Bianchi identity
by integrating both sides of expression~\eqref{BI} over closed
5-cycles. Since the components of the Bianchi identity have a
$y$ index, one of the dimensions of the cycle must be the orbifold
direction. The remaining dimensions form some closed four-cycle in the Calabi-Yau
threefold. The left-hand side of~\eqref{BI}, being exact, gives zero upon
integration, whereas each term on the right-hand side becomes a
topological invariant.  We thus obtain the well-known cohomology
condition of heterotic M-theory for each choice of four-cycle
${\cal C}_4$, \bea \label{cohcond} -\frac{1}{16 \pi^2} \int_{{\cal
    C}_4} \tr F^{(1)} \wedge F^{(1)} - \frac{1}{16 \pi^2} \int_{{\cal
    C}_4} \tr F^{(2)} \wedge F^{(2)} + \frac{1}{16 \pi^2} \int_{{\cal
    C}_4} \tr R \wedge R + \sum_{p=1}^N \int_{{\cal C}_4} J^{(p)} =0 \;.
\eea The physical interpretation of this condition is simply that the
sum of the charges on the compact space must vanish since there is nowhere
for the ``field lines'' to go.

One can expand the charges \eqref{charges} in terms of the eigenmodes
of the Laplacian on the $CY_3$ \cite{Lukas:1998hk}.  Consider the
two-forms satisfying the eigenmode equation \bea \triangle
\omega_{<m>\;AB} = - \lambda_{<m>}^2 \omega_{<m>\;AB} \;, \eea where
$\triangle$ is the Laplacian and the index $<m>$ labels the eigenmode
on the threefold with eigenvalue $-\lambda^{2}_{<m>}$.  Note that
eigenmodes with different values of $<m>$ may have identical
eigenvalues.  For example, consider the number of two-form eigenmodes
with $\lambda^{2}_{<m>}=0$, that is, the zero modes of
$\triangle$. Since by definition these two-forms are harmonic, there
are ${\rm dim}H^2(X)={\rm dim}H^{(1,1)}$ of them. We will denote these
harmonic modes by $\omega_{k}$, where $k=1, \dots , h^{1,1}$. Note
that eigenmodes corresponding to a non-vanishing eigenvalue
$-\lambda^{2}_{<m>} \neq 0$ are neither harmonic nor, in general,
closed forms. The metric on the space of eigenmodes, \bea
G_{<m><n>}=\frac{1}{2v}\int_{CY_3}{\omega_{<m>} \wedge
  \ast_{CY_3}\omega_{<n>}}
\label{metric1}
\eea where $\ast_{CY_3}$ is the Poincar\'e duality operator on the
$CY_3$, is used to raise and lower $<m>$-type indices.

Each of the four-form charges $J^{(p)}$, $p=0,...,N+1$ in~\eqref{charges}
can then be expanded as \bea \label{modeexp} \ast_{CY_3} J^{(p)} = \frac{1}{2
 v^{\frac{2}{3}}} \sum_{<m>} \beta_{<m>}^{(p)} \omega^{<m>} \;.\eea 
 Using metric~\eqref{metric1}, it follows that 
 \bea
  \beta_{<m>}^{(p)}=\frac{1}{v^{1/3}}\int_{CY_{3}}{\omega_{<m>} \wedge J^{(p)}}.
  \label{q1}
  \eea
 In particular, for the coefficients $\beta_{k}^{(p)}$ corresponding to the zero modes one finds
 \bea
 \beta_{k}^{(p)}=\int _{{\cal{C}}_{4k}}{J^{(p)}},
 \label{q2}
 \eea
 where ${\cal{C}}_{4k}$, $k=1, \dots , h^{1,1}$ are the four-cycles dual to the harmonic $(1,1)$ forms $\omega_{k}$. It then follows from the the cohomology
condition \eqref{cohcond} that
\bea
\label{zeromodesum} \sum_{p=0}^{N+1} \beta_k^{(p)} =0 
\eea 
for each integer $k$. In general, as
indicated in \eqref{modeexp}, the four-form charges in \eqref{BI} are
built out of both harmonic and non-harmonic modes on the $CY_3$. There
is no similar condition on the sum of the coefficients of the
non-harmonic components of the charges.

The closed four-form charges in heterotic M-theory have a definite index
structure in terms of the complex structure of $CY_3$. The gauge
field in a supersymmetric compactification is obtained by solving the
killing spinor equations for the gauginos - the so-called Hermitian
Yang-Mills equations \cite{Green:1987mn}. This results in a field
strength which is a $(1,1)$ form. Thus $F \wedge F$ is a $(2,2)$
form. For the Ricci flat metric on a $CY_3$, $R \wedge R$ is similarly
a $(2,2)$ form. Finally, solving the killing spinor equations for the
worldvolume fermions on the bulk five-branes tells us that
these object wrap holomorphic curves. This implies that the $J^{(p)}$
four-form charges of the bulk branes are also $(2,2)$ forms, via
the definition given in, and underneath, equation \eqref{charges}. It follows that, written in terms of complex indices, Bianchi identity~\eqref{BI} decomposes into two conditions given by
\begin{equation}
(dG)_{ya\bar{b}c\bar{d}}=4 \pi
\left(\frac{\kappa_{11}}{4 \pi} \right)^{\frac{2}{3}} \left[ \delta
  (y) J^{(0)} + \delta (y - \pi \rho) J^{(N+1)} + \frac{1}{2}
  \sum_{p=1}^{N} J^{(p)} \left( \delta (y - y_{(p)}) + \delta (y +
    y_{(p)}) \right) \right]_{a\bar{b}c\bar{d}}
    \label{BI1}
\end{equation}
and
 \begin{equation}
 (dG)_{abc\bar{d}y} =0
 \label{BI2}
 \end{equation}
and its Hermitian conjugate respectively.

In summary, in complex coordinates we have two components of the Bianchi identity~\eqref{BI}. The first of these,~\eqref{BI1}, contains closed $(2,2)$
form charges which are, in general, composed of a series of harmonic and
non-harmonic modes on the $CY_3$. The coefficients of the harmonic
modes in the expansion of these charges sum to zero, as shown in
equation~\eqref{zeromodesum}. The remaining component of the Bianchi identity, given in~\eqref{BI2}, has vanishing sources. We now turn to the solutions to these two constraints.

\subsection{Solving for the background flux}
\label{zeromodesonly}

\subsubsection{The non-zero mode}
\label{nzm}

Let us begin by considering the first constraint,
equation~\eqref{BI1}. Expanding the left hand side in components
yields
$\partial_{y}G_{a\bar{b}c\bar{d}}+2\partial_{[a}G_{|\bar{b}|c]\bar{d}y}-
2\partial_{[\bar{b}}G_{|ac|\bar{d}]y}$. Recall from subsection 3.2
that the source terms in the Bianchi identity all involve delta
functions with $y$ in the argument. It follows that the sources on the
right hand side of~\eqref{BI1} can only be obtained from a
non-singular, if discontinuous, $G$ as a $y$-derivative of
$G_{a\bar{b}c\bar{d}}$. We now solve for the $(2,2)$ component of $G$
that saturates these charges.

The standard approach to finding this component is to restrict attention
to the harmonic modes in the expansion of the charges
\eqref{modeexp}. This is justified by the fact \cite{Lukas:1998hk}
that the effects of the non-harmonic modes are suppressed relative to
those of the harmonic ones by powers of the parameter $\epsilon_R$.
A solution to the Bianchi identity \eqref{BI1} in this approximation is
given by having a non-zero expectation value for $G_{a \bar{b} c
  \bar{d}}$ which jumps at each extended object by an amount
proportional to that objects charge and is  constant in $y$ everywhere else.
These requirements are satisfied by
\bea
\label{nonzeromode} G^{(nzm)} _{a\bar{b}c\bar{d}}= -\frac{1}{V^{\frac{1}{3}}}
\sum_{k=1}^{h^{1,1}} \alpha_k(y) ( \ast_{CY_3} \omega_k)_{\a\bar{b}c\bar{d}} \;,
\eea where 
\bea \alpha_k (y) = \frac{2 \pi}{v^{\frac{2}{3}}}
\left( \frac{\kappa_{11}}{4 \pi}\right)^{\frac{2}{3}}\left[
  \beta_k^{(0)} \theta (y) + \beta_k^{(N+1)} \theta (y - \pi \rho) +
  \sum_{p=1}^N \beta^{(p)}_k \left( \theta(y-y_{(p)}) +\theta
    (y+y_{(p)}) \right) \right]\;. 
\label{alpha1}    
    \eea 
Taking the $y$-derivative of this expression exactly reproduces the 
charges on the right hand side of ~\eqref{BI1} in the harmonic mode approximation. 
Note that $G^{(nzm)} _{a\bar{b}c\bar{d}}$ in~\eqref{nonzeromode} is a closed
$(2,2)$ form on the $CY_{3}$.

Now consider the $(1,2)$ and $(2,1)$ components of $G$,
$G_{\bar{b}c\bar{d}y}$ and $G_{ac\bar{d}y}$ respectively, within the
harmonic approximation.  For values of $y$ in the bulk space, that is,
away from the extended sources, identity~\eqref{BI1} becomes
\begin{equation}
\partial_{y}G^{(nzm)}_{a\bar{b}c\bar{d}}+2\partial_{[a}G_{|\bar{b}|c]\bar{d}y}-
2\partial_{[\bar{b}}G_{|ac|\bar{d}]y}=0.
\label{again1}
\end{equation}
Since $G^{(nzm)}_{a\bar{b}c\bar{d}}$ in~\eqref{nonzeromode} is constant between the sources, it follows that $\partial_{y}G^{(nzm)}_{a\bar{b}c\bar{d}}=0$ in~\eqref{again1} and, hence,
\begin{equation}
2 \partial_{[a}G_{|\bar{b}|c]\bar{d}y} - 2 \partial_{[\bar{b}}G_{|ac|\bar{d}]y} \equiv (d^{CY_3} G)_{a \bar{b} c \bar{d} y}=0,
\label{again2}
\end{equation}
where $d^{CY_3}$ is the exterior derivative operator on the $CY_3$.
Non-trivial solutions to equation \eqref{again2} are both possible and
important, and we will discuss them in detail in the following
subsection.  Here, however, we simply note that, in the harmonic
approximation to the sources, the non-vanishing component
$G^{(nzm)}_{a\bar{b}c\bar{d}}$ is sufficient to satisfy the
identity~\eqref{BI1} everywhere in the orbifold interval, both at each
extended source and in the bulk space.

We now examine the second constraint equation, given
in~\eqref{BI2}. In components, this becomes
\begin{equation}
  \partial_{y}G_{abc\bar{d}}+3 \partial_{[a}G_{bc]\bar{d}y}-\partial_{\bar{d}}G_{abcy}=0.
\label{again3}
\end{equation}
As discussed previously, the $(3,1)$ component $G_{abc\bar{d}}$ is odd
under the ${\mathbb Z}_{2}$ orbifolding. It follows that, if
non-vanishing, this component would require source charges on the
right hand side of~\eqref{again3}. Since none exist, we must set
\begin{equation}
G_{abc\bar{d}}=0.
\label{again4}
\end{equation}
Equation~\eqref{again3} then becomes
\begin{equation}
  3 \partial_{[a}G_{bc]\bar{d}y} - \partial_{\bar{d}}G_{abcy} \equiv (d^{CY_3} G)_{abc \bar{d} y}=0.
\label{again5}
\end{equation}
Combining \eqref{again5} with \eqref{again2} and their complex
conjugates, we see that the $(2,1)$ and $(3,0)$ components of $G$,
$G_{bc\bar{d}y}$ and $G_{abcy}$ respectively, and their conjugates are
the components of a closed three-form on the $CY_{3}$. Again, these
closed forms are important, and will be discussed in detail in the
next subsection. Here, however, we simply note that setting them to
zero is sufficient to satisfy the Bianchi identity~\eqref{BI1} and
\eqref{BI2} .

$G^{(nzm)}_{a\bar{b}c\bar{d}}$ in equation \eqref{nonzeromode} is the
famous non-zero mode of heterotic M-theory - hence our choice of
labeling superscript. The fact that, in the harmonic approximation,
there is no $G_{ABCy}$ component to the expectation value, together
with the lack of bulk $y$ dependence of the non-zero mode, means that
if we compactify the eleven-dimensional theory down to five-dimensions
we will obtain an action in the bulk, between any two extended
objects, which does not depend explicitly on $y$. This structure is
the basis for the very existence of 5D heterotic
M-theory~\cite{Lukas:1998tt}.  In the low energy limit, heterotic
$M$-theory further reduces to an effective, four-dimensional,
${\cal{N}}=1$ supersymmetric theory. It is well-known that the
harmonic part of the non-zero mode does {\it not} give rise to a
potential in four dimensions. This has been shown many times by
explicit dimensional reduction. The 4D superpotential contribution due
to flux in heterotic M-theory takes the Gukov-Vafa-Witten form \bea
  \label{GVW} W_{GVW} \propto \int_{{\cal M}_7} \Omega \wedge G. \eea
 Here $\Omega$ is the holomorphic three-form on the $CY_3$. It is
 obvious that if we substitute $G^{(nzm)}_{a\bar{b}c\bar{d}}$ into this expression we
 get zero, since neither the flux nor $\Omega$ carry a $y$ index. Note that this will continue to be the case for $G^{(nzm)}_{a\bar{b}c\bar{d}}$ even when non-harmonic modes are included.

\vspace{0.1cm}

Thus far, we have solved for the non-zero mode of heterotic M-theory in the approximation that the 
charges $*_{CY_{3}}J^{(p)}$,$p=0,\dots,N+1$ in~\eqref{BI1} are harmonic $(2,2)$ forms.
In the remainder of this subsection, we generalize these results to include the effects of the non-harmonic contributions to the charges. To balance the delta function on
the right hand side of \eqref{BI1}, one still requires the field $G$ to jump
as it crosses each extended object in the $y$ direction. Even
including the non-harmonic components of the charges, all of the
four-forms on the right hand side of \eqref{BI1} have a $(2,2)$ index
structure. Hence, it is the $(2,2)$ component of the four-form field
strength which must jump at the extended sources, as it did in the harmonic approximation~\eqref{nonzeromode}. However, recall that while the sum of the harmonic contributions
to the charges is zero, equation~\eqref{zeromodesum}, the same is {\it not} true
for the non-harmonic modes. Because of this, when the non-harmonic modes are included,
it no longer suffices to simply have the
$(2,2)$ component of $G$ jump at the extended objects and be constant
everywhere else. It was shown in~\cite{Lukas:1998hk} that to globally obey all of the boundary
conditions, the $G^{(nzm)}_{a\bar{b}c\bar{d}}$ component of the four-form has to evolve in $y$ in the
bulk space so that it may undergo the correct jump at each charged
object.

This observation has implications for the $(1,2)$ and $(2,1)$ forms,
$G_{\bar{b}c\bar{d}y}$ and $G_{ac\bar{d}y}$. In between the extended
sources, identity~\eqref{BI1} is given by
equation~\eqref{again1}. Clearly, if $G^{(nzm)}_{a\bar{b}c\bar{d}}$ is
a non-constant function of $y$ in the bulk, we require non-vanishing
$G_{\bar{b}c\bar{d}y}$ and $G_{ac\bar{d}y}$ components which are {\it
  not} closed on the $CY_{3}$. These components of $G$, which we denote by
$G^{(nzm)}_{\bar{b}c\bar{d}y}$ and $G^{(nzm)}_{ac\bar{d}y}$
respectively, correspond in the ten-dimensional theory to the $H$ flux
resulting from a non-standard embedding. The details of the solution
of the Bianchi identity~\eqref{BI1} and the equations of motion for
the case where non-harmonic modes are included can be found in
\cite{Lukas:1998hk}. The explicit functional form of
$G^{(nzm)}_{a\bar{b}c\bar{d}}$, $G^{(nzm)}_{\bar{b}c\bar{d}y}$ and
$G^{(nzm)}_{ac\bar{d}y}$, including their exact $y$-dependence, is
given in that paper. The expressions are somewhat large and, since no
additional information beyond the comments made above are required in
this paper, we will not reproduce them here.

Can the new $G^{(nzm)}_{\bar{b}c\bar{d}y}$ and
$G^{(nzm)}_{ac\bar{d}y}$ components of the non-zero mode now give rise
to a contribution to the heterotic $M$-theory 4D superpotential:
$W_{GVW} = \int_{{\cal M}_7} \Omega \wedge G$?  Since these forms have
a $y$ index, they can at least saturate the interval part of the
integral. However, recall that these components are $(1,2)$ and
$(2,1)$ forms on the $CY_{3}$ respectively.  Since $\Omega$ is a
holomorphic $(3,0)$ form, such flux can not give a non-vanishing
contribution to the Gukov-Vafa-Witten expression for the
superpotential. In addition to the argument presented here, based upon
previous microscopic derivations of the superpotential, valid to some
order in the expansions of heterotic M-theory, there are a variety of
macroscopic arguments based upon non-renormalization theorems of
\cite{Witten:1985bz,Dine:1986vd,Green:1987mn}. These say that the
non-zero mode, including its non-harmonic pieces, should never
contribute to the superpotential of supersymmetric heterotic M-theory
at any order.

\subsubsection{Harmonic flux}
\label{harmflux}

The $G^{(nzm)}_{a\bar{b}c\bar{d}}$ and $G^{(nzm)}_{\bar{b}c\bar{d}y}$,
$G^{(nzm)}_{ac\bar{d}y}$ forms comprising the non-zero mode are not
the most general solution to the Bianchi
identity~\eqref{BI1},~\eqref{BI2} and the equations of motion. Recall
that the components of $G^{(nzm)}$ with a $y$ index, when evaluated
using both harmonic and non-harmonic contributions to the charges, are
{\it not} closed on the $CY_{3}$. As discussed above, one may add to $G$
any {\it closed} forms with index structure
\begin{equation}
G_{a\bar{b}\bar{c}y}, G_{\bar{a}bcy}, G_{abcy}, G_{\bar{a}\bar{b}\bar{c}y},
\label{bell1}
\end{equation}
that is, any element of $\textnormal{dim}H^{3}(X) =
\textnormal{dim}H^{1,2} \oplus \textnormal{dim} H^{2,1} \oplus
\textnormal{dim}H^{3,0} \oplus \textnormal{dim}H^{0,3}$, and still
satisfy the Bianchi identity. By choosing these forms to be not just
closed but also harmonic in seven dimensions, they continue to satisfy
the equations of motion. For this to be the case, one must choose the
harmonic representative in each $CY_{3}$ cohomology class and restrict
these to be constant in $y$.  We denote this harmonic flux
contribution to $G$ by $G^{H}$. That is, the components of $G^{H}$ are
the $y$-independent harmonic representatives of the
$2(h^{2,1}+1)$-dimensional cohomology space $H^{3}(X)$. It is these
closed forms that we will refer to as the flux contributions to
heterotic vacua.

This $G^{H}$ contribution to $G$ {\it does} give rise to a
superpotential in the four-dimensional theory; specifically, for the
complex structure moduli of the $CY_3$. We will reproduce the
derivation of this, and present, for the first time, its extension to
the case where anti-branes are present, in the next section. For now,
we simply note that the harmonic flux includes a $(0,3)$ form
component $G_{\bar{a} \bar{b} \bar{c}y}$, which can give a non-zero
contribution to a superpotential $W_{GVW} \propto \int_{{\cal M}_7}
\Omega \wedge G$ when combined with the holomorphic $(3,0)$ form
$\Omega$. The reader should not think that the above index structure
considerations imply that only one of the $2(h^{2,1} +1)$ flux
parameters, that corresponding to the $H^{0,3}$ component, contributes
to the superpotential. The superpotential depends, in general, on the
complex structure moduli. As these change their values, the component
of the flux which corresponds to the ``$(0,3)$'' piece also
changes. For example, as is well known and will be shown again below,
in the large complex structure limit \bea \label{lcs}
W_{\textnormal{flux}} = \frac{\sqrt{2}}{\kappa_4^2} \epsilon_0
\frac{v^{\frac{1}{6}}}{(\pi \rho)^2} \left( \frac{1}{6}
  \tilde{d}_{\underline{a} \underline{b} \underline{c}}
  \mathfrak{z}^{\underline{a}} \mathfrak{z}^{\underline{b}}
  \mathfrak{z}^{\underline{c}} n^0 - \frac{1}{2}
  \tilde{d}_{\underline{a} \underline{b} \underline{c}}
  \mathfrak{z}^{\underline{a}} \mathfrak{z}^{\underline{b}}
  n^{\underline{c}} - \mathfrak{z}^{\underline{a}} n_{\underline{a}} -
  n_0 \right)\; .  \eea Here $n^0$, $n_0$, $n^{\underline{a}}$ and
$n_{\underline{a}}$ are the flux parameters and the $\tilde{d}$ are
the intersection numbers on the mirror $CY_3$. For any given value of
the complex structure fields $\mathfrak{z}^{\underline{a}}$, we obtain
a superpotential depending upon a single combination of the
parameters. However, as the $\mathfrak{z}^{\underline{a}}$ change, the
combination of parameters which appears in the above expression also
changes. The superpotential then, as a function on field space,
depends on all $2(h^{2,1}+1)$ parameters.  Henceforth, we will label
the $G^{H}$ contribution to Gukov-Vafa-Witten 4D superpotential as
$W_{\rm flux}$.

\vspace{0.1cm}

\subsubsection{Flux quantization and the derivation of the 4d potential.}
\label{quant}

To summarize: within the context of supersymmetric heterotic M-theory
we have considered two contributions to the vacuum expectation value
of $G$. The first is the non-zero mode $G^{(nzm)}_{a\bar{b}c\bar{d}}$
and $G^{(nzm)}_{\bar{b}c\bar{d}y}$,$G^{(nzm)}_{ac\bar{d}y}$; that is,
the form-flux sourced by the extended objects in the theory. The
non-zero mode is, in general, composed of both harmonic and
non-harmonic modes on the $CY_3$. The second contribution to $G$ we
refer to as the harmonic flux $G^{H}$. This is a `free field'
background flux which is composed exclusively of the $2(h^{2,1}+1)$
harmonic forms $G_{a\bar{b}\bar{c}y}, G_{\bar{a}bcy}, G_{abcy},
G_{\bar{a}\bar{b}\bar{c}y}$ on the internal manifold.  Of these two
contributions to $G$, only the harmonic flux gives rise to a
superpotential, and so a potential, in four dimensions. In this
section, we show that the $2(h^{2,1}+1)$ forms in $G^{H}$ are
quantized and derive the four dimensional potential which the harmonic
flux gives rise to.

As shown in \cite{Rohm:1985jv,Witten:1996md} and generalized here, the four-form
vacuum expectation value $G$ obeys a quantization condition derived by
demanding that the supermembrane path integral be well-defined in the
background under consideration. This condition is found to be
\bea \label{QC}  \left( \frac{4 \pi}{\kappa_{11}}
\right)^{\frac{2}{3}} \int_{{\cal C}_4} \frac{G}{2 \pi} + \int_{{\cal
    C}_4} \frac{\lambda}{2} + \frac{1}{16 \pi^2} \sum_{i=1}^{2}
\int_{\partial {\cal C}_4^{(i)}} \omega^{\textnormal{YM}(i)} -
\sum_{p=1}^N \int_{{\cal C}_4} \omega_3^{(p)} (\delta (y- y^{(p)})) =
n\; , \eea where ${\cal{C}}_{4}$ is any four-cycle in ${\cal{M}}_{7}$
and $n$ is an arbitrary integer. The four-form $\lambda$ is half of the first
Pontryagin class of the compactification manifold. It is associated
with the form $\frac{1}{16\pi^{2}} R \wedge R$ and locally can be
written as $\lambda=d\omega^{L}$, where $\omega^{L}$ is the Lorentz
Chern-Simon three-form. Similarly, $\omega^{YM(i)}$ is the Yang-Mills
Chern-Simon term on the $i$-th orbifold fixed plane and the three-form
$\omega^{(p)}_3$ is defined locally by $d \omega^{(p)}_3 = J^{(p)}$ on
the $p$-th M5 brane.

We begin by by choosing the cycle ${\cal{C}}_{4}$ to lie entirely in the $CY_{3}$. Since such a cycle has no boundary in the Calabi-Yau threefold and no component in the orbifold $y$ direction, it follows that condition~\eqref{QC} simplifies to
\bea \label{QC1}  \left( \frac{4 \pi}{\kappa_{11}}
\right)^{\frac{2}{3}} \int_{{\cal C}_4} \frac{G}{2 \pi} + \int_{{\cal
    C}_4} \frac{\lambda}{2}  =
n\;.  \eea 
Recall that the $G^{(nzm)}_{a\bar{b}c\bar{d}}$ component of the non-zero mode is the only background component of $G$ with all indices in the $CY_{3}$. Fixing the Pontryagin class of the $CY_{3}$, we conclude that $G^{(nzm)}_{a\bar{b}c\bar{d}}$ is  quantized as in~\eqref{QC1}.

Now choose the cycle ${\cal{C}}_{4}$ in~\eqref{QC} to be composed of a
three-cycle in the $CY_{3}$ and a component in the $y$ direction. Clearly
this singles out the
$G^{(nzm)}_{\bar{b}c\bar{d}y}$ and $G^{(nzm)}_{ac\bar{d}y}$ components of
the non-zero mode, as well as the harmonic forms
$G_{a\bar{b}\bar{c}y}, G_{\bar{a}bcy}, G_{abcy}$ and
$G_{\bar{a}\bar{b}\bar{c}y}$. First consider the non-zero mode
$G^{(nzm)}_{\bar{b}c\bar{d}y}$, $G^{(nzm)}_{ac\bar{d}y}$. As was shown
in~\cite{Witten:1996md,Lukas:1997rb}, Bianchi identity~\eqref{BI}
guarantees that these two components satisfy \bea \label{QC2}
 \left( \frac{4 \pi}{\kappa_{11}}
\right)^{\frac{2}{3}} \int_{{\cal C}_4} \frac{G}{2 \pi} = -\int_{{\cal
    C}_4} \frac{\lambda}{2} - \frac{1}{16 \pi^2} \sum_{i=1}^{2}
\int_{\partial {\cal C}_4^{(i)}} \omega^{\textnormal{YM}(i)} +
\sum_{p=1}^N \int_{{\cal C}_4} \omega_3^{(p)} (\delta (y- y^{(p)}))\;.
\eea That is, for fixed background $CY_{3}$ and Yang-Mills gauge
connection the non-zero mode components $G^{(nzm)}_{\bar{b}c\bar{d}y}$
and $G^{(nzm)}_{ac\bar{d}y}$ are determined and cancel out of
quantization condition~\eqref{QC}. This condition now simplifies to
\bea \label{11fluxquant}  \left( \frac{4
    \pi}{\kappa_{11}} \right)^{\frac{2}{3}} \int_{{\cal C}_4}
\frac{G^H}{2 \pi} = n\;. \eea
 
A more explicit expression for the quantization of the harmonic flux
$G^{H}$ can be found by expanding its components~\eqref{bell1} in
terms of the a basis of the associated cohomology
groups. Specifically, let $\alpha_{\underline{A}}$ and
$\beta^{\underline{B}}$ be a basis of the cohomology $H^{2,1} \oplus
H^{3,0}$ and $H^{1,2} \oplus H^{0,3}$ of the Calabi-Yau
threefold. Written, for simplicity, in terms of the real indices on the
$CY_{3}$, one has \bea \label{11dfluxes} G_{ABCy} = {\cal
  X}^{\underline{A }}_{y}\alpha_{{\underline{A}} ABC} - \tilde{{\cal
    X}}_{y \underline{B}} \beta^{\underline{B}}_{ABC}.  \eea By
definition, as was discussed explicitly in~\cite{mattias}, these
coefficients are related to the fields $\xi^{\underline{A}}$ and
$\tilde{\xi}_{\underline{B}}$ in the five-dimensional theory through
the $y$-component of their field strengths \bea \label{5dfluxes1}
{\cal X}^{\underline{A}} \equiv {\cal X}^{\underline{A}}_y
= \partial_y \xi^{\underline{A}}, \;\; \tilde{{\cal
    X}}_{\underline{B}} \equiv \tilde{{\cal X}}_{{\underline{B}} y}
= \partial_y \tilde{\xi}_{\underline{B}}. \eea In our application, we
must take \bea \label{5dfluxes2} {\cal X}^{\underline{A}}=
\textnormal{constant} \; , \;\; \tilde{{\cal X}}_{\underline{B}}=
\textnormal{constant} \eea in order for the associated $G_{ABCy}$ to
be harmonic in the seven-dimensional sense.

Inserting expression~\eqref{11dfluxes} into condition~\eqref{11fluxquant} gives rise to the quantization of the $2(h^{2,1}+1)$ constants in equation \eqref{5dfluxes2} when one integrates over
the appropriate cycles. 
Let us define the three-cycles $a^{\underline{A}}$ and
$b_{\underline{B}}$ in the Calabi-Yau threefold by \bea \label{cycledef}
\int_{X}\a_{\underline{B}}\w\b^{\underline{A}}=v^{\frac{1}{2}} \int_{a^{\underline{A}}}\a_{\underline{B}}= v \; \d_{\underline{B}}^{\underline{A}}\; ,\quad
\int_{X}\b^{\underline{A}}\w\a_{\underline{B}}=v^{\frac{1}{2}} \int_{b_{\underline{B}}}\b^{\underline{A}}=- v \; \d_{\underline{B}}^{\underline{A}} \; , \quad \int_{a^{\underline{A}}} \beta^{\underline{B}} = 0 \; , \quad \int_{b_{\underline{B}}} \alpha^{\underline{A}} =0. \eea They form
a basis for the homology $H_{1,2} \oplus H_{0,3}$ and $H_{2,1} \oplus
H_{3,0}$  which is dual to $\alpha_{\underline{A}}$ and
$\beta^{\underline{B}}$. First consider a four-cycle ${\cal C}_4$ composed of
three-cycle $a^{\underline{A}}$ and the orbifold direction.  When the harmonic
component of the flux given in equation \eqref{11dfluxes} is
integrated over this four-cycle, condition \eqref{11fluxquant}
becomes, using~\eqref{cycledef}, \bea
\left( \frac{4 \pi}{\kappa_{11}} \right)^{\frac{2}{3}} \int_{a^{\underline{A}}}
\frac{{\cal X}^{\underline{B}} \alpha_{\underline{B}} \pi \rho}{2 \pi} = n^{\underline{A}}. \eea Referring to
\eqref{cycledef} again to perform the final integral, as well as to the definitions in equations \eqref{BB} and \eqref{AA}, we find that
\bea \label{4dfluxes1} \frac{1}{ \epsilon_0}  \frac{ (\pi
  \rho)^2}{v^{\frac{1}{6}}} {\cal X}^{\underline{A}} = n^{\underline{A}} \;. \eea Here the $n^{\underline{A}}$'s are arbitrary integers
for each $\underline{A}$.
A similar calculation, where the $a^{\underline{A}}$ part of the
four-cycle is replaced by $b_{\underline{B}}$, demonstrates that the $\tilde{{\cal
    X}}$'s are quantized in a similar manner. That is,
\bea \label{4dfluxes2}  \frac{1}{ \epsilon_0} \frac{ (\pi
  \rho)^2}{v^{\frac{1}{6}}} \tilde{{\cal X}}_{\underline{B}} = n_{\underline{B}}\;, \eea
where the $n_{\underline{B}}$ are arbitrary integers for each $\underline{B}$.

\vspace{0.2cm}

Let us now consider the dimensional reduction from the five- to the four-dimensional
supersymmetric theory including this quantized flux, using the methods introduced in~\cite{paper1}. The starting point is the five-dimensional action given in~\eqref{5daction} with $N$ M5-branes but {\it no anti M5-branes}.  We see from \eqref{5daction} that there are only two
terms in the five-dimensional action which contain the scalar fields
$\xi^{\underline{A}}$ and $\tilde{\xi}_{\underline{B}}$. These are \bea \label{fluxaction5d}
&&-\frac{1}{2 \kappa_5^2} \int d^5 x \sqrt{-g} \left[ - V^{-1} (
  \tilde{{\cal X}}_{\alpha {\underline{B}}} - \bar{M}_{{\underline{B}}{\underline{C}}} ({\mathfrak z}) {\cal
    X}_{\alpha}^{\underline{C}}) ( [ \textnormal{Im}{(M ({\mathfrak
      z}))}]^{-1})^{{\underline{B}}{\underline{A}}} ( \tilde{{\cal X}}_{\alpha {\underline{A}}} - M_{{\underline{A}}{\underline{D}}}
  ({\mathfrak z}) {\cal X}_{\alpha}^{\underline{D}}) \right] \\ \nonumber &&
\;\;\;\;\;\;\quad\quad \quad \quad \quad \quad \quad \quad \quad \quad
\quad - \frac{1}{2 \kappa_5^2} \int \left( 2 G \wedge ( \xi^{\underline{A}}
  \tilde{{\cal X}}_{\underline{A}} - \tilde{\xi}_{\underline{A}} {\cal X}^{\underline{A}}) \right). \eea To find
the terms in the four-dimensional theory which depend on the fluxes,
therefore, it suffices to study the dimensional reduction of these two
terms. The quantization conditions \eqref{4dfluxes1} and
\eqref{4dfluxes2} show that the fluxes themselves are already first
order in the $\kappa_{11}^{\frac{2}{3}}$ expansion. This makes the
first term in \eqref{fluxaction5d} second order in this expansion. In
counting these orders, one should ignore, as usual
\cite{Horava:1996ma}, the overall prefactor of the action proportional
to $\kappa_{11}^{-2}$.  Explicit calculation reveals that the terms
involving flux in the second component of \eqref{fluxaction5d} will
also be at least second order in $\kappa_{11}^{\frac{2}{3}}$ and, even
at this order, contain at least one four-dimensional derivative. Since
we are interested here in obtaining potential energy terms only, we
discard these contributions henceforth.

A straight forward dimensional reduction of the first term of equation
\eqref{fluxaction5d}, using the reduction ansatz described
in~\cite{Brandle:2001ts,paper1}, gives the following as the
flux-dependent contribution to the potential energy in
four-dimensions.  \bea \label{fullpotential} -\frac{1}{2 \kappa_4^2}
\int d^4 x \sqrt{-g_4} \left[ - \frac{1}{b_{0}^{3}V_0} (\tilde{{\cal
      X}}_{\underline{A}} - \bar{M}_{{\underline{A}}{\underline{B}}}
  ({\mathfrak z}) {\cal X}^{\underline{B}}) ( [ \textnormal{Im}{(M
    ({\mathfrak z}))}]^{-1})^{{\underline{A}}{\underline{C}}} (
  \tilde{{\cal X}}_{\underline{C}} -
  M_{{\underline{C}}{\underline{D}}} ({\mathfrak z}) {\cal
    X}^{\underline{D}}) \right]. \eea Here $\kappa_{4}$ is the
four-dimensional Planck constant defined by
$\kappa^{2}_{4}=\kappa^{2}_{5}/\pi\rho$, and ${\cal X}$, $\tilde{{\cal
    X}}$ are the quantized quantities described in \eqref{4dfluxes1}
and \eqref{4dfluxes2}. An examination of the parameters appearing in
this action, and the quantized quantities therein, reveals that these
terms are of order $\epsilon_S^2 \epsilon_R^2$. The fact that they are
already second order in the strong coupling parameter $\epsilon_S$,
means that we will not consider higher order corrections to this flux
potential arising from the warping. Such contributions are small and
beyond the order to which we calculate the four-dimensional effective
theory in this paper.

It is well-known that flux potentials in supersymmetric theories
should be derivable from a Gukov-Vafa-Witten type superpotential of
the form \bea \label{Wflux} W_{\textnormal{flux}} =
\frac{\sqrt{2}}{\kappa_4^2} \frac{1}{\pi \rho \, v^{\frac{1}{2}}}
\int_{X \times S^1/\mathbb{Z}_2} \Omega \wedge G \;. \eea
Substituting~\eqref{11dfluxes} into this expression gives
\bea \label{Wflux2} W_{\textnormal{flux}} =
\frac{\sqrt{2}}{\kappa_4^2} 
\left( {\cal X}^{\underline{A}} {\cal G}_{\underline{A}} -
  \tilde{{\cal X}}_{\underline{B}} {\cal Z}^{\underline{B}} \right)
\;, \eea where the complex structure moduli space is parametrized by
the periods $({\cal Z}^{\underline{B}}, {\cal G}_{\underline{A}}({\cal
  Z}))$ defined as \footnote{Note that the ${\cal Z}^{\underline{B}}$
  denote a set of projective coordinates on the complex structure
  moduli space. One can obtain a set of affine coordinates by the
  usual procedure of picking one non-vanishing homogeneous coordinate
  and dividing the others with respect to it: that is, ${\mathfrak
    z}^{\underline{a}} = {\cal Z}^{\underline{a}}/{\cal Z}^0$,
  $a=1,\dots,h^{2,1}$ when ${\cal Z}^0$ is not zero. It is this set of
  affine coordinates that appears in action \eqref{5daction}.}
\begin{equation}
{\cal Z}^{\underline{B}}=\int_{a^{\underline{B}}}{\Omega}, \quad {\cal G}_{\underline{A}}({\cal Z})=\int_{b_{\underline{A}}}{\Omega}.
\label{more}
\end{equation}

It is now necessary to show that the term~\eqref{fullpotential} in the four-dimensional action is actually of this form.  Applying the usual
supergravity formalism, using the K\"ahler potential found in
\cite{paper1} and reproduced in \eqref{modk} of Appendix A, we find that
this is indeed the case. However, we postpone a proof of this until the next section and Appendix B, where we also include anti M5-branes.
Here, instead, we will assume that~\eqref{Wflux} is the correct four-dimensional flux superpotential and use~\eqref{Wflux2} and~\eqref{more} to calculate its explicit form in terms of the affine complex coordinates ${\mathfrak z}^{\underline{a}}$ in the large complex structure limit. 

We proceed as follows. It is a well known fact (following from an
examination of the large K\"ahler modulus limit of the mirror
compactification) that the prepotential, ${\cal G}$, takes the
following form in the large complex structure limit.
\begin{eqnarray} {\cal G} = - \frac{1}{6}
  \frac{\tilde{d}_{\underline{a} \underline{b} \underline{c}} {\cal
      Z}^{\underline{a}} {\cal Z}^{\underline{b}} {\cal
      Z}^{\underline{c}}}{{\cal Z}^0} \end{eqnarray} Note here we are
splitting up the index $\underline{A}$ into an index $\underline{a}$
and the remaining possible value $0$. Substituting this expression
into \eqref{Wflux2} we find the following.
\begin{eqnarray}
  W_{\textnormal{flux}} = \frac{\sqrt{2}}{\kappa_4^2} \left( \frac{1}{6} \frac{\tilde{d}_{\underline{a}
        \underline{b} \underline{c}} {\cal Z}^{\underline{a}} {\cal
        Z}^{\underline{b}} {\cal Z}^{\underline{c}}}{({\cal Z}^0)^2} {\cal X}^0 - \frac{1}{2} \frac{\tilde{d}_{\underline{a}
        \underline{b} \underline{c}} {\cal Z}^{\underline{a}} {\cal
        Z}^{\underline{b}}}{{\cal Z}^0} {\cal X}^{\underline{c}} - \tilde{{\cal X}}_{\underline{a}} {\cal Z}^{\underline{a}}   - \tilde{{\cal X}}_0 {\cal Z}^0 \right)
\end{eqnarray}
We now use the definition of the affine coordinates,
$\mathfrak{z}^{\underline{a}} = \frac{{\cal Z}^{\underline{a}}}{{\cal
    Z}^0}$. We also use the scale invariance of the physical theory
under rescalings of the homogeneous coordinates ${\cal Z}$ to set
${\cal Z}^0$ to $1$.
\begin{eqnarray}
  W_{\textnormal{flux}} = \frac{\sqrt{2}}{\kappa_4^2} \left( \frac{1}{6} \tilde{d}_{\underline{a} \underline{b} \underline{c}} \mathfrak{z}^{\underline{a}} \mathfrak{z}^{\underline{b}} \mathfrak{z}^{\underline{c}} {\cal X}^0 - \frac{1}{2} \tilde{d}_{\underline{a} \underline{b} \underline{c}} \mathfrak{z}^{\underline{a}} \mathfrak{z}^{\underline{b}} {\cal X}^{\underline{c}} - \tilde{{\cal X}}_{\underline{a}} \mathfrak{z}^{\underline{a}} -  \tilde{{\cal X}}_0 \right)
\end{eqnarray}
Using equations \eqref{4dfluxes1} and \eqref{4dfluxes2} we obtain,
finally, \eqref{lcs} which we repeat here.
\begin{eqnarray}
  W_{\textnormal{flux}} = \frac{\sqrt{2}}{\kappa_4^2} \epsilon_0 \frac{v^{\frac{1}{6}}}{(\pi \rho)^2} \left( \frac{1}{6} \tilde{d}_{\underline{a} \underline{b} \underline{c}} \mathfrak{z}^{\underline{a}} \mathfrak{z}^{\underline{b}} \mathfrak{z}^{\underline{c}} n^0 - \frac{1}{2} \tilde{d}_{\underline{a} \underline{b} \underline{c}} \mathfrak{z}^{\underline{a}} \mathfrak{z}^{\underline{b}} n^{\underline{c}} -  \mathfrak{z}^{\underline{a}} n_{\underline{a}} -  n_0 \right)
\end{eqnarray}

\vspace{0.2cm}

The results in this section were derived implicitly assuming two
important constraints on the magnitude of the $G$-form expectation
values. The first of these concerns the value of the non-zero
mode, $G^{(nzm)}_{a\bar{b}c\bar{d}}$. The smallness of the $\epsilon_S$ and $\epsilon_R$
parameters ensures that the back-reaction of this flux is adequately
described by the warping it induces along the orbifold
direction. Hence, one can continue to perform the analysis on a
Calabi-Yau threefold, despite the presence of this component of the
background flux, although the geometry of this space does change along
the orbifold direction. This assumption is standard in the discussion
of all strong coupling heterotic vacua.  The second assumption
concerns the magnitude of the $G$-flux corresponding to the various non-vanishing
$G_{ABCy}$ harmonic components. For these components, we are making
the standard ``weak flux'' approximation. That is, we assume that,
despite the presence of these $G$-fluxes, one can still compactify on
a Calabi-Yau threefold and do not require a more general manifold of
$SU(3)$ structure. This approximation is particularly easy to control
from the point of view of the five-dimensional theory. The fluxes are
expectation values of the $y$-derivative of certain five-dimensional
moduli. Thus, the Calabi-Yau approximation is valid whenever the
five-dimensional $y$-derivatives of the associated moduli are small
compared to the Calabi-Yau compactification scale. This will be the
case whenever the number of units of flux is chosen to be sufficiently
small and the moduli take appropriate values.

Finally, one may ask about the effect of the diffuse source of
curvature, which the flux represents, on the bulk warping. The above
discussion makes it clear that, for the case where we have a small
number of units of flux quanta, this warping modification can only
come in at second order in $\epsilon_S$. This, as was explained in
\cite{paper1}, is at a higher order than is needed to calculate the
action of the theory to the orders we consider here. Therefore, we can
consistently neglected this contribution to the warping \footnote{It
  is interesting to note that this correction to the warping is of the
  same size as the potential term we are keeping. This is consistent
  because of the normalization we choose for our moduli fields. This
  normalization ensures that the correction to the action, which comes
  from terms linear in this warping, vanishes and only the quadratic
  and higher contributions are present.}.
  
Having completed our review of flux in supersymmetric Heterotic
M-theory let us now proceed to examine how the above discussion
changes in the presence of anti-branes.

 \section{Flux in Heterotic M-Theory with Anti-Branes}
\label{fluxsection} 
  
In this section, the explicit contributions of flux to the four-dimensional effective action of heterotic M-theory {\it with anti M5-branes} will be derived. As in the previous section, we find it most transparent to begin the discussion in the eleven-dimensional context. The role of flux in five and four dimensions is then derived by dimensional reduction. The exposition is similar to that given in the supersymmetric case. Hence, we use Section 3 as a template, explicitly showing how anti M5-branes alter the conclusions therein. As discussed in Section 2, we will assume there are $N-1$ M5-branes and a single anti M5-brane indexed by $(\bar{p})$ in the bulk space. The extension of these results to an arbitrary number of anti branes is straightforward.

To begin, note that the discussion of the components of background four-form flux $G$ that are relevant to stable vacua with maximal spacetime symmetry, Subsection 3.1, is unchanged by the addition of anti branes. However, an anti M5-brane located in the bulk space does alter
the Bianchi identity and its sources discussed in Subsection 3.2. Consider Bianchi identity~\eqref{BI}. The right hand side is sourced by the four-form charges localized on the orbifold fixed planes and $N$ M5-branes given in~\eqref{charges}. In the case of $N-1$ M5-branes and a single anti M5-brane, the form of expression~\eqref{BI} remains unchanged. However, the sum on the right hand side now includes an anti M5-brane charge $J^{(\bar{p})}$. The charges on the two orbifold planes and the $N-1$ M5-branes given in~\eqref{charges} remain unchanged. However, the anti M5-brane four-form charge is given by
\begin{equation}
J^{(\bar{p})}=-\delta({\cal{C}}^{(\bar{p})}_{2}).
\label{fun1}
\end{equation}
Note that ${\cal{C}}^{(\bar{p})}_{2}$ remains a holomorphic curve on which the anti M5-brane is wrapped, the reverse orientation of the brane being expressed by the minus sign. As with all the other four-forms in~\eqref{charges},  $J^{(\bar{p})}$ is closed on the $CY_{3}$. With the proviso that 
$J^{(\bar{p})}$ be given by~\eqref{fun1}  rather than the last line in~\eqref{charges}, every expression and conclusion of Subsection 3.2 remains unchanged.

\subsection{The non-zero mode}

Now reconsider Subsection 3.3, where we solve for the background flux,
in the presence of an anti M5-brane. Let us begin with the non-zero
mode discussed in Subsection 3.3.1. As above, all expressions and
equations in this subsection remain unchanged with the proviso that
$J^{(\bar{p})}$ be given by~\eqref{fun1} everywhere. Be that as it
may, the physical conclusions for the four-dimensional theory {\it
  change dramatically}. To see this, first consider the harmonic
approximation to the non-zero mode given
in~\eqref{nonzeromode},~\eqref{alpha1}, where, now,
$\bar{\beta}_{k}=\beta^{(\bar{p})}_k$ is defined by
expressions~\eqref{q2} and~\eqref{fun1}. It was stated
in~\cite{paper1} that, when the eleven-dimensional theory is
dimensionally reduced on the $CY_{3}$ in the background of this
non-zero mode, the effective five-dimensional theory is given by the
action in equation~\eqref{5daction}. As discussed in Section 2, the
$N-1$ M5-branes and the anti M5-brane contributions to the non-zero
mode enter action~\eqref{5daction} through the coefficients
$\beta^{(p)}_{k}$ which appear in $\hat{\beta}_{k}$, $n^{k}_{(p)}$ and
$j_{(p)\mu}$. It is important to note that despite the appearance of
an anti M5-brane in the vacuum, action~\eqref{5daction} is
${\cal{N}}=1$ supersymmetric, both in the five-dimensional bulk space
and on all four-dimensional fixed planes and branes. The reason for
this is that the dimensional reduction was carried out with respect to
the supersymmetry preserving Calabi-Yau threefold, and did not
explicitly involve the anti M5-brane. However, as discussed
in~\cite{paper1} and Section 2 of this paper, the dimensional
reduction to four-dimensions {\it does} involve the anti
M5-brane. Hence, one expects the four-dimensional effective action to
explicitly break ${\cal{N}}=1$ supersymmetry. This is indeed the case,
as was shown in~\cite{paper1}. Remarkably, the effect of the anti
brane was found to appear in only two specific places in the action,
to the order at which we calculate. First, the four-dimensional theory
exhibits a potential energy for the moduli. Secondly, the kinetic
energy functions for the gauge fields, specifically, the coupling of
their field strengths to moduli, is modified to include
non-holomorphic terms. Both of these effects explicitly break
supersymmetry and both vanish if the anti M5-brane is removed from the
theory. In this section, we will discuss the first of these, that is,
the supersymmetry breaking potential energy. We defer the discussion
of the modified gauge couplings to the following section on gaugino
condensation, where it becomes relevant.

The explicit form of the supersymmetry breaking moduli potential energy for an arbitrary number of K\"ahler and complex structure moduli, $N-1$ M5-branes and one anti M5-brane was given in Section 5 of~\cite{paper1}.  The potential, written in terms of the complex scalar components of the moduli superfields,  was found to be 
\begin{equation}
{\cal V} = {\cal V}_1+{\cal V}_2,
\label{india7}
\end{equation}
where
\begin{equation}
{\cal{V}}_{1} = \frac{\epsilon_0 \, \kappa_4^{-2}}{(\pi \rho)^2}
(T^k+\bar{T}^k) \delta_k e^{\kappa_4^2 (K_T+K_D)}   
\label{india8}
\end{equation}
and
\bea \label{india9} {\cal{V}}_{2}= \kappa_4^{-2} \frac{\epsilon_0^2}{(\pi \rho)^2}
e^{\kappa_4^2 (K_T +2 K_D)} K_T^{\bar{k} l} \delta_l \left[ \sum_{p=0}^{\bar{p}-1}
  \tau^{(p)}_k \frac{Z_{(\bar{p})} + \bar{Z}_{(\bar{p})}}{\bar{\tau}_m
    (T^m +\bar{T}^m)} - \sum_{p=\bar{p}+1}^{N+1} \tau^{(p)}_k
  \frac{Z_{(\bar{p})} + \bar{Z}_{(\bar{p})}}{\bar{\tau}_m (T^m
    +\bar{T}^m)} \right. \\ \left. \nonumber - \sum_{p=0}^{\bar{p}-1}
  \tau^{(p)}_k \frac{Z_{(p)} + \bar{Z}_{(p)}}{\tau_m^{(p)} (T^m
    +\bar{T}^m)} + \sum_{p=\bar{p}+1}^{N+1} \tau^{(p)}_k \frac{Z_{(p)}
    + \bar{Z}_{(p)}}{\tau_m^{(p)} (T^m +\bar{T}^m)} \right. \\
\left. \nonumber + \sum_{p=0}^{N+1} \tau^{(p)}_k \left( 1-
    \frac{Z_{(p)}+\bar{Z}_{(p)}}{\tau^{(p)}_m (T^m+\bar{T}^m)} \right)
  \frac{Z_{(p)}+\bar{Z}_{(p)}}{\tau^{(p)}_n (T^n+\bar{T}^n)}
  -\frac{2}{3} \delta_k \right]
\eea
are the $\kappa^{\frac{2}{3}}_{11}$ and  $\kappa^{\frac{4}{3}}_{11}$ contributions respectively.
The complex scalar fields $T^{k}$, $k=1,\dots,h^{1,1}$ and $Z_{(p)}$, $p=0,\dots,\bar{p},\dots,N+1$, corresponding to the K\"ahler moduli and the location moduli of the M5-branes and anti M5-brane respectively, are defined, along with the complex dilaton scalar $S$, in equation~\eqref{superfields} of Appendix A. Similarly, the K\"ahler potentials $K_T$ and $K_D$ are given 
in~\eqref{K} of that Appendix. They key point is that both contributions~\eqref{india8} and~\eqref{india9} to ${\cal{V}}$ are proportional to the anti M5-brane tension
\begin{equation}
\delta_{k}=\frac{1}{2}({\bar{\tau}}_{k}-\bar{\beta}_{k})={\bar{\tau}}_{k},
\label{india10}
\end{equation}
where we have used the notation, introduced earlier, that
$\bar{\tau_{k}}=\tau^{(\bar{p})}_{k}$ and
$\bar{\beta}_{k}=\beta^{(\bar{p})}_{k}$. Hence, the non-vanishing of
this potential is due entirely to the existence of the anti
M5-brane. We conclude that when sourced by an anti M5-brane, the
harmonic contribution to the non-zero mode
$G^{(nzm)}_{a\bar{b}c\bar{d}}$, together with a series of other
contributions such as the sum of the tensions of the extended objects,
{\it does} induce a non-vanishing supersymmetry breaking potential
energy in the four-dimensional theory.

Were the anti M5-brane to be removed and replaced by an M5-brane, all $\delta_{k}\rightarrow 0$ and, hence, ${\cal{V}}$ would vanish. This result is completely consistent with the statement given in Subsection 3.3.1 that in the supersymmetric case with no anti M5-brane, the non-zero mode component $G^{(nzm)}_{a\bar{b}c\bar{d}}$ cannot contribute to the Gukov-Vafa-Witten superpotential~\eqref{GVW} and, hence, leads to vanishing potential energy for the moduli fields in the four-dimensional theory. The vanishing of~\eqref{india7} in the case of no anti M5-brane and, hence, $\delta_{k}\rightarrow 0$, constitutes an explicit proof by dimensional reduction of this conclusion.
Note, however, that when an anti M5-brane is present, the potential energy is non-supersymmetric and need not be derived from a superpotential. 

Let us now include the non-harmonic contributions to the four-form charges of the orbifold planes, the $N-1$ M5-branes and the anti M5-brane. The effect of this is to add an additional contribution to the 
$G^{(nzm)}_{a\bar{b}c\bar{d}}$ component of the non-zero mode. Furthermore, as discussed in Subsection 3.3.1, the non-harmonic modes will induce non-vanishing values for the $(1,2)$ and $(2,1)$
components, $G^{(nzm)}_{\bar{b}c\bar{d}y}$ and $G^{(nzm)}_{ac\bar{d}y}$ respectively. One expects these additional non-harmonic contributions to the non-zero mode to induce corrections to the four-dimensional supersymmetry breaking potential~\eqref{india7}. This is indeed the case. However, as discussed in~\cite{Lukas:1998hk} such additional contributions are suppressed relative to the harmonic contributions by powers of the parameter $\epsilon_{R}$. For that reason, we do not display their explicit form here. Suffice it to say that these terms remain proportional to the anti M5-brane tension $\delta_{k}={\bar{\tau}}_{k}$. Once again, if the anti M5-brane is removed and replaced with an M5-brane, then $\delta_{k}\rightarrow 0$ and these  terms vanish. Again, this is completely consistent with the statement in Subsection 3.3.1 that neither $G^{(nzm)}_{\bar{b}c\bar{d}y}$ nor $G^{(nzm)}_{ac\bar{d}y}$ can contribute to the Gukov-Vafa-Witten superpotential~\eqref{GVW}.

\subsection{Harmonic flux and flux quantization}

Having discussed the non-zero mode, let us now reconsider Subsection
3.3.2 and Subsection 3.3.3, the harmonic flux and its flux
quantization and contribution to the 4d effective action respectively,
in the presence of an anti M5-brane. This is easy to do since,
remarkably, the anti brane does not alter any of the conclusions of
those subsections. First, consider Subsection 3.3.2. Since they are
closed on the $CY_{3}$, the $2(h^{2,1}+1)$ harmonic forms
$G_{a\bar{b}\bar{c}y}, G_{\bar{a}bcy}, G_{abcy},
G_{\bar{a}\bar{b}\bar{c}y}$ in $G^{H}$ are not sourced by the fixed
planes, $N-1$ M5-branes or the anti M5-brane. Hence, the addition of
the anti M5-brane does not effect the harmonic flux in any way. Nor
does it effect the conclusions about the Gukov-Vafa-Witten
superpotential and expression~\eqref{lcs} in that subsection. Since
supersymmetry {\it is} broken in the four-dimensional effective theory
by the anti M5-brane, this last statement requires further discussion,
which we will return to shortly.

Now consider Subsection 3.3.3. First note that the flux quantization
condition~\eqref{QC}, as well as the constraint equation~\eqref{QC2}
for the non-zero mode components $G^{(nzm)}_{\bar{b}c\bar{d}y}$ and
$G^{(nzm)}_{ac\bar{d}y}$, remain unchanged in the presence of an anti
M5-brane, with the proviso that the three-form
$\omega^{(\bar{p})}_{3}$ associated with the anti brane is defined
locally by $d\omega^{(\bar{p})}_{3}=J^{(\bar{p})}$ where
$J^{(\bar{p})}$ is given by~\eqref{fun1}. It follows that the
quantization condition~\eqref{QC1} for the non-zero mode
$G^{(nzm)}_{a\bar{b}c\bar{d}}$ and the quantization
condition~\eqref{11fluxquant} for the harmonic modes $G^{H}$ are also
unchanged. Furthermore, the presence of the anti M5-brane does not
alter the definitions of the four-dimensional flux constants ${\cal
  X}^{\underline{A}}, \tilde{{\cal X}}_{\underline{B}}$ or their
quantization conditions given in~\eqref{4dfluxes1}
and~\eqref{4dfluxes2} respectively.

One must now consider the dimensional reduction of the five- to the four-dimensional theory including this quantized flux. Here, however, this must be carried out in the presence of an anti M5-brane. Again, the starting point is the five-dimensional action given in~\eqref{5daction}, now, however, 
with $N-1$ M5-branes and an anti M5-brane. As discussed earlier, the form of this action does not change when an anti brane is included in the vacuum. Hence, the two terms in the five-dimensional action containing the scalar fields $\xi^{\underline{A}}$ and $\tilde{\xi}_{\underline{B}}$ are still given by expression~\eqref{fluxaction5d}. As discussed previously, only the first term in this expression can contribute to the potential energy. We now perform a dimensional reduction of the first term in~\eqref{fluxaction5d} with respect to the supersymmetry breaking vacuum described in~\cite{paper1} and Section 2 of this paper. Remarkably, despite the fact that this background contains an anti M5-brane,
we find that the flux-dependent contribution to the potential energy in four-dimensions is given by
\bea \label{fullpotential2} -\frac{1}{2 \kappa_4^2} \int d^4 x
\sqrt{-g_4} \left[ - \frac{1}{b_{0}^{3}V_0} (\tilde{{\cal X}}_{\underline{A}} -
  \bar{M}_{{\underline{A}}{\underline{B}}} ({\mathfrak z}) {\cal X}^{\underline{B}}) ( [ \textnormal{Im}{(M
    ({\mathfrak z}))}]^{-1})^{{\underline{A}}{\underline{C}}} ( \tilde{{\cal X}}_{\underline{C}} - M_{{\underline{C}}{\underline{D}}}
  ({\mathfrak z}) {\cal X}^{\underline{D}}) \right], \eea
that is, the same expression~\eqref{fullpotential} as in the supersymmetric case, up to the order in our expansions which we work to. Hence, although the anti M5-brane does give rise to explicit  supersymmetry breaking terms in the four-dimensional theory, the flux sector of the effective theory remains ${\cal{N}}=1$ supersymmetric. 

Since this term appears in a four-dimensional ${\cal{N}}=1$ supersymmetric action (in the situation without anti-branes), it must be possible to express the Lagrangian density of~\eqref{fullpotential2} in the form
\begin{equation}
V_{\rm flux}=e^{\kappa^{2}_{4}K_{\rm mod}}\left(K^{i\bar{j}}_{\textnormal{mod}} 
D_{i}W_{\rm flux}\overline{D_{j}W_{\rm flux}}-3\k_{4}^{2}|W_{\rm flux}|^{2}\right)\,,
\label{wow1}
\end{equation}
where $K_{\rm mod}$ is the K\"ahler potential of the moduli and $W_{\rm flux}$ is the holomorphic superpotential generated by the harmonic flux. This is indeed the case. The proof, as originally given in \cite{mattias}, is somewhat intricate, so we relegate it to Appendix B. The result is that the Lagrangian density of~\eqref{fullpotential2} can be written in the form~\eqref{wow1} where the K\"ahler potential $K_{\rm mod}$ is given in expression~\eqref{modk} of Appendix A and $W_{\rm flux}$ is the Gukov-Vafa-Witten superpotential
\bea \label{Wflux3} W_{\textnormal{flux}} =
\frac{\sqrt{2}}{\kappa_4^2} \frac{1}{ \pi \rho \, v^{\frac{1}{2}}} \int_{X \times S^1/\mathbb{Z}_2} \Omega
\wedge G =
\frac{\sqrt{2}}{\kappa_4^2}
\left( {\cal X}^{\underline{A}} {\cal G}_{\underline{A}} - \tilde{{\cal X}}_{\underline{B}} {\cal Z}^{\underline{B}} \right) \;, \eea where the complex structure moduli space is
parametrized by the periods $({\cal Z}^{\underline{B}}, {\cal G}_{\underline{A}}({\cal Z}))$
defined by \begin{equation}
{\cal Z}^{\underline{B}}=\int_{a^{\underline{B}}}{\Omega}, \quad {\cal G}_{\underline{A}}({\cal Z})=\int_{b_{\underline{A}}}{\Omega}.
\label{amore}
\end{equation}
This conclusion is valid for both the supersymmetric case and when there is an anti M5-brane in the vacuum. It is consistent with, and proves, the form of the flux superpotential presented in~\eqref{Wflux} and~\eqref{Wflux2} at the end of Subsection 3.3.3. Similarly, the expression given in Subsection 3.3.3 for the superpotential in the large complex structure limit,
\bea \label{lcsagain} W_{\textnormal{flux}} = \frac{\sqrt{2}}{\kappa_4^2} \epsilon_0 \frac{v^{\frac{1}{6}}}{(\pi \rho)^2} \left( \frac{1}{6} \tilde{d}_{\underline{a} \underline{b} \underline{c}} \mathfrak{z}^{\underline{a}} \mathfrak{z}^{\underline{b}} \mathfrak{z}^{\underline{c}} n^0 - \frac{1}{2} \tilde{d}_{\underline{a} \underline{b} \underline{c}} \mathfrak{z}^{\underline{a}} \mathfrak{z}^{\underline{b}} n^{\underline{c}} - \mathfrak{z}^{\underline{a}} n_{\underline{a}} -  n_0 \right) \; ,  \eea
remains unchanged in the presence of an anti M5-brane.

\section{Gaugino Condensation in Heterotic M-Theory with
  Anti-Branes}
\label{sectiongaugino}

In this section, we discuss gaugino condensation in the presence
of anti-branes. Our exposition will proceed in several steps. First,
we show that the form of the potential energy terms arising from gaugino
condensation, when written in terms of the condensate itself, are
unchanged from the result one obtains without
anti-branes. Second, we argue that a gaugino condensate 
will indeed occur in this non-supersymmetric
setting, despite the fact that most of our knowledge of this effect is
based on the structure of unbroken $N=1$ supersymmetric gauge theory. 
Finally, we compute the condensate as an explicit
function of the moduli fields. We conclude that the significant changes that
anti-branes induce in the gauge kinetic functions of the orbifold
gauge fields lead, via the gaugino condensate,  to important 
modifications of the potential energy.

\subsection{The potential as a function of the condensate}

We begin by reviewing how gaugino condensates are included in
the dimensional reduction of five-dimensional heterotic M-theory, as
was first presented in \cite{Lukas:1999kt}. We then show that, when written in 
terms of the condensate itself, the presence of anti-branes does not alter the form of
the potential. Unlike the flux contribution to the potential energy, which is most naturally described in terms of the $2(h^{2,1}+1)$ real scalar fields $\xi^{\underline{A}}$ and ${\tilde{\xi}}_{\underline{B}}$, the gaugino condensate
contribution is most simply expressed in terms of the flux of a single complex scalar field
$\xi$. These fields are related by
\bea \label{redefA}
\left(\ba{rr} \xi^{\underline{A}} \\ \tilde{\xi}_{\underline{B}}\ea \right) = \left( \ba{cc}
  {\cal Z}^{\underline{A}}
 & f^{\underline{A}}_{\underline{a}} \\ {\cal G}_{\underline{B}}& h_{\underline{a}\underline{B}} \ea \right) \left(\ba{rr} \xi \\ \eta^{\underline{a}} \ea \right) + h.c., \eea
where the periods $({\cal{Z}}^{\underline{A}}, {\cal{G}}_{\underline{B}})$ were defined in~\eqref{more} and the expressions for 
$f^{\underline{A}}_{\underline{a}} $ and $h_{\underline{a}\underline{B}} $ are found in the Appendix of~\cite{mattias}. The field $\xi$ supplies two of the four bosonic components of the ``universal'' hypermultiplet in five-dimensions, whereas the complex scalar fields $\eta^{\underline{a}}$ are half of the bosonic components of the remaining $h^{2,1}$ hypermultiplets.
The relevant quantity for gaugino condensation is ${\cal X}_{\alpha}=\partial_{\alpha}\xi$, the five-dimensional field strength associated with $\xi$.

As obtained by dimensional reduction from eleven-dimensions,  the five-dimensional 
action of heterotic M-theory, where the fields $\xi^{\underline{A}}$, ${\tilde{\xi}}_{\underline{B}}$ are written in terms of $\xi$, $ \eta^{\underline{a}}$ using~\eqref{redefA}, contains a term, consisting of $\xi$-flux and gaugino 
bilinears , which is a ``complete square''.
This specific term is familiar from both the eleven-dimensional theory and the ten-dimensional 
weakly coupled heterotic string \cite{Dine:1985rz,Horava:1996vs}. 
The first step in including gaugino
condensates in the reduction of the five-dimensional theory is to
define a new set of fields. This avoids the appearance of squares of
delta functions in the discussion.  In five dimensions, the relevant
field redefinitions are 
\bea
X_{\mu} &=& {\cal X}_{\mu}, \\
X_{y} &=& {\cal X}_y + \frac{1}{32 \pi}
\frac{\kappa_5^2}{\alpha_{\textnormal{GUT}}} \left[\Lambda^{(0)}
  \delta ( y) + \Lambda^{(N+1)} \delta (y -\pi \rho) \right], \eea 
where the quantities $\Lambda^{(0)}$ and $\Lambda^{(N+1)}$
are the condensates, that is, the fermion bilinears, themselves. For completeness, we have
introduced a condensate on each orbifold plane. One of these can always be set to
zero, if so required.\footnote{It should be noted that the gauge
  groups which appear on various stacks of branes and anti-branes
  in the bulk could also give rise to condensation in situations where
  they are strongly coupled and non-Abelian. We will ignore this
  possibility here, although it should be kept in mind in
  a detailed discussion of moduli stabilization.}

In terms of these new quantities, the complete square in the
five-dimensional action simply becomes a kinetic term for
$X_{\alpha}$, as can be seen using equation
\eqref{5daction}. Therefore, the only place where the condensate
explicitly appears is in the Bianchi identity of $X_{\alpha}$. This is given by~\cite{Lukas:1999kt} 
\bea (dX)_{y \mu} = -\frac{\kappa_5^2}{32 \pi
  \alpha_{\textnormal{GUT}}} \left[(4 J^{(0)}_{\mu} + \partial_{\mu}
  \Lambda^{(0)}) \delta(y) + (4 J^{(N+1)}_{\mu} + \partial_{\mu}
  \Lambda^{(N+1)}) \delta(y- \pi \rho) \right] ,
  \label{who}
  \eea
 where $J_{\mu}^{(i)}$ is proportional to $\partial_{\mu}W_{(i)}$, the derivative of the matter field superpotential on the $i$-th orbifold plane.
Note that, in deriving \eqref{who}, it is essential to realize that the condensate
can depend on the four-dimensional moduli. Otherwise,
one would have $\partial_{\mu}\Lambda^{(i)}=0$ and the source terms in the Bianchi identity
would be condensate independent. This would lead to vital terms in the dimensional
reduction being missed.
It is also important to note that the anti-branes do not appear in 
Bianchi identity \eqref{who}. As was discussed in \cite{paper1}, anti-branes
simply do not source the bulk fields involved in the present
discussion. This is one of the essential features of anti-branes.
It ensures that the potential due to flux and gaugino condensation
is of the same form when anti-branes are present as it is in the supersymmetric case.

To take into account the presence of a gaugino condensate, we simply
have to solve the above system and then perform a dimensional
reduction about the result. A solution for $X_{\alpha}$ has, in
fact, several contributions. There is one contribution
induced by matter field fluctuations on the boundaries, as indicated
in the above by the source terms $J_{\mu}^{(i)}$ in the Bianchi
identity. Next there is a contribution due to whatever harmonic flux we may
choose to turn on, as described in the previous section. Finally,
there is the contribution obtained from the magnetic charges for
this field proportional to the derivative of the condensate. Since all
of these contributions are small in our approximations, we can treat
each of them separately, simply adding the results to obtain the full
expression for $X_{\alpha}$.  In this section, we discuss the last of these, that is, 
the contribution to $X_{\alpha}$ due to the condensate. This contribution will be 
denoted by $X_{\alpha}^{\Lambda}$.  We find that
\bea \label{whoa1} X^{\Lambda}_y &=& \frac{\kappa_5^2}{64
  \pi^2 \rho \, \alpha_{\textnormal{GUT}}} \left(
  \Lambda^{(0)} + \Lambda^{(N+1)}\right), \\
X^{\Lambda}_{\mu} &=& -\frac{\kappa_5^2}{64 \pi
  \alpha_{\textnormal{GUT}}} \partial_{\mu} \left( \Lambda^{(0)} -
  \frac{y}{\pi \rho} (\Lambda^{(0)}+ \Lambda^{(N+1)} )\right).
  \label{whoa}
  \eea

  Having found the complex $\xi$-flux induced by gaugino condensation,
  we would like to re-express it in the same field basis as in
  the previous section, that is, in terms of the flux ${\cal{X}}^{\underline{A}}$
  and ${\tilde{\cal{X}}}_{\underline{B}}$ of the real scalar fields $\xi^{\underline{A}}$ and
  ${\tilde{\xi}}_{\underline{B}}$. This is easily done using
  relation~\eqref{redefA}. Differentiating this relation, we find
  that the gaugino condensate makes the following contributions to
  these fluxes.  \bea \label{wow}{\cal X}^{{\underline{A}} \Lambda} &=& {\cal Z}^{\underline{A}}
  X^{\Lambda}_y
  + \bar{{\cal Z}}^{\underline{A}} \bar{X}^{\Lambda}_y \nonumber \\
  \tilde{{\cal X}}_{\underline{B}}^{\Lambda} &=& {\cal G}_{\underline{B}} X^{\Lambda}_y +
  \bar{{\cal G}}_{\underline{B}} \bar{X}^{\Lambda}_y.
  \eea This result uses the fact
  that the warping in the complex structure moduli is first order in
  our expansions. Therefore, since the condensate contribution to the $\xi$-flux is already small,
 this warping can be neglected here.

  Let us insert this contribution to the ${\cal X}$ flux, along with the 
  contributions discussed in the previous section, into \eqref{fullpotential2}. This gives us the
  four-dimensional potential energy 
  due to both gaugino condensation and harmonic flux. The result is 
  \bea -\frac{1}{2 \kappa_4^2}&& \int d^4 x
  \sqrt{-g_4} \left[ - \frac{1}{b_{0}^{3}V_0} \left(
      (\tilde{{\cal X}}_{\underline{A}} + \tilde{{\cal X}}_{\underline{A}}^{\Lambda}) -
      \bar{M}_{\underline{A}\underline{B}} ({\mathfrak z}) ({\cal X}^{\underline{B}} + {\cal X}^{\underline{B} \Lambda})
    \right) ( [ \textnormal{Im}{(M ({\mathfrak z}))}]^{-1})^{\underline{A}\underline{C}}
  \right. \\ \nonumber && \left.  \times \left( (\tilde{{\cal X}}_{\underline{C}}+
      \tilde{{\cal X}}_{\underline{C}}^{\Lambda}) - M_{\underline{C}\underline{D}} ({\mathfrak z}) ({\cal
        X}^{\underline{D}}+ {\cal X}^{\underline{D} \Lambda}) \right) \right].  \eea Note that,
  as in the previous section, the corrections to the warping of the
  bulk fields due to the presence of anti-branes do not change the form of the 
  above expression from the supersymmetric result. This is due to $1)$ the
  order at which the above terms appear in the expansion parameters
  of heterotic M-theory and $2)$ the low scale of the condensate. These
  considerations make such corrections outside of the approximations
  to which we work.

This result can be processed into a more user friendly form by
employing the results on special geometry given in Appendix B. Using~\eqref{wow},
identities \eqref{ident1} and \eqref{ident2}, the fact that $M(
{\mathfrak z})$ is a symmetric matrix and the definitions of the flux
superpotential and K\"ahler potentials given in \eqref{Wflux3} and~\eqref{modk},~\eqref{K} respectively, we find that the potential energy is
\bea \label{componentresult} 
V_{\rm c+c/f+f} = \frac{1}{2 \kappa_4^2}
e^{\kappa_{4}^{2}(K_{D}+K_{T})}\left( 2 e^{ - \kappa_4^2 {\cal K}({ \mathfrak z})} |
    X^{\Lambda}_y |^2 + i X^{\Lambda}_y
       \sqrt{2} \kappa_4^2 W_{\textnormal{flux}} - i \sqrt{2} \kappa_4^2
   \bar{W}_{\textnormal{flux}} \bar{ X}^{\Lambda}_y
 \right) 
  \nonumber 
 \eea
 
 \bea
 + e^{ \kappa_4^2 K_{\rm mod}} \left( K_{\rm mod}^{i
 \bar{j}} D_i W_{\textnormal{flux}} D_{\bar{j}}
\bar{W}_{\textnormal{flux}} - 3 \kappa_4^2
|W_{\textnormal{flux}}|^2 \right)  .
 \eea

Note that, as in the pure flux potential discussed in the previous section, the pure 
gaugino condensation and flux-gaugino condensation cross-term contributions to the potential for the moduli are independent of both the M5-branes and anti M5-branes in the vacuum, up to
second order in $\epsilon_S$ and the small condensate scale. Hence,
the functional form of expression~\eqref{componentresult} is the same
whether or not anti-branes are present.

How, then, does the action for theories with and without anti-branes
differ. The answer lies in the explicit form of the condensates
$\Lambda^{(0)}$ and $\Lambda^{(N+1)}$ when expressed in terms of the
moduli fields. It follows from~\eqref{whoa1} that these condensates
determine $X^{\Lambda}_y$, which we must now
specify. Before doing this, first notice from~\eqref{whoa1} that
$X^{\Lambda}_y $ is proportional to the sum of the two
condensates. It turns out that one can calculate $X^{\Lambda}_y $
for each condensate separately, simply adding the results at the end
of the computation. We can, therefore, restrict the discussion to a
single boundary wall.

\subsection{Condensate scales I: supersymmetric case}

In this subsection, we review the computation of the condensate $X^{\Lambda}_y $
for the supersymmetric case without anti-branes. 
Let the gauge bundle in this sector have structure group $G$.
Then the low energy theory contains a super Yang-Mills connection with structure group $H$, where $H$ is the commutant of $G$ in $E_{8}$. For simplicity, we will assume that $H$ is a simple Lie group. This Yang-Mills theory is coupled in the usual way to supergravity. Therefore, the gauge coupling is determined by the values of certain moduli. There can also be matter multiplets in this sector. However, they are irrelevant to the discussion in this paper and, with the exception of their contribution to the beta-function,  we will ignore them. Such a theory is
known to undergo gaugino condensation under certain conditions
\cite{Veneziano:1982ah}.  

Gaugino condensation induces a superpotential for the moduli fields
entering the gauge coupling. This superpotential has been calculated,
in \cite{Taylor:1990wr} for example, using an effective low energy
field theory where the condensate itself is the lowest component of a
superfield. For the simple case where the gauge coupling depends on
the dilaton modulus $S$ only, this superpotential is found to be
\bea \label{wgaugino} W_{\textnormal{gaugino}} = A e^{-\epsilon S} \;,
\eea where, at tree level, $A$ is a constant of order
$v^{-\frac{1}{2}}$,
\begin{equation}
\epsilon=\frac{6\pi}{{\bf b}_{0}\alpha_{GUT}}
\label{whoa2}
\end{equation}
and ${\bf b}_{0}$ is the beta function coefficient for the effective gauge theory. The dilaton field $S$ is defined in our context in~\eqref{superfields} of Appendix A. For example, for $H=E_{8}$ there are no matter fields and
\begin{equation}
{\bf b}_{0}=90.
\label{whoa3}
\end{equation}

Adding $W_{\textnormal{gaugino}}$ to $W_{\textnormal{flux}}$, and
inserting them into the usual supergravity formalism, yields the
complete moduli potential energy induced by gaugino condensation and
flux. To evaluate the condensate $X^{\Lambda}_y $, one need only
compare this potential energy to
expression~\eqref{componentresult}. This is most easily done at low
energy and for large values of the moduli. We will refer to this as
the ``gaugino condensation limit''. In this limit, we find that the
leading terms for the pure gaugino and gaugino/flux contributions to
the supergravity potential simplify to the following result,
\bea \label{potential1} V_{\rm c+c/f} = e^{\kappa_4^2 K_{\rm mod}}
\left[ \epsilon^2 (S+\bar{S})^2 \kappa_4^2 |A|^2 e^{-\epsilon
    (S+\bar{S})} + \kappa_4^2 \epsilon (S + \bar{S}) (A e^{-\epsilon
    S} \bar{W}_{flux} + \textnormal{h.c}) \right] \;.\eea This
expression is precisely reproduced by equation \eqref{componentresult}
if one chooses the leading order contribution to the condensate to be
\bea \label{condensate}{\bar{ X}}^{\Lambda}_y = i e^{{\kappa_4^2 \cal
    K} ({\mathfrak z})} \frac{{A}}{\sqrt{2}} \kappa_4^2 \epsilon
(S+\bar{S}) e^{-\epsilon S}. \eea Note that it is perfectly consistent
for the chiral condensate, which is the lowest component of a chiral
superfield in the effective theory, to take a non-holomorphic
form. This is because this relation simply describes the value of the
field in vacuum and is not an identity on field space.

Expression~\eqref{condensate} was computed in the gaugino condensation
limit. Can one find an expression for $X^{\Lambda}_y $ at any energy
and for arbitrary moduli expectation values? To do this, the complete
supergravity moduli potential induced by $W_{\textnormal{gaugino}}$
and $W_{\textnormal{flux}}$ must be compared to
expression~\eqref{componentresult} and the functional form of the
condensate inferred. One finds that the condensate continues to be proportional to the same exponential factor of the gauge kinetic function as in the gaugino condensation limit. However, 
in general, the prefactor is now a {\it very} complicated function of the moduli.

That it is difficult to reproduce the lower order terms in the large
moduli expectation values, as was first noted within the context of
the weakly coupled heterotic string in \cite{Dine:1985rz}, should not
come as a surprise. First, parts of this potential correspond to
gravitational corrections to the global supersymmetry result.  One
should, therefore, also introduce gravitational corrections to the
value of the condensate once we include these terms. Gravitationally,
of course, the condensate is coupled to everything in the theory and
so it will take a very complicated form at this level. Second,
component analyses of the kind performed in the previous subsection
and \cite{Dine:1985rz} are somewhat naive. We should also include the
effects of integrating out the strongly coupled non-Abelian gauge
fields and so forth. This could also introduce new terms which would
contribute to the problematic polynomial prefactor.

From a physical point of view, the inability to reproduce the
polynomial prefactor for arbitrary moduli expectation values is not
very important. In any regime of moduli space which is well described
by effective supergravity, the real parts of the moduli fields, in
particular the dilaton and K\"ahler moduli, should be much greater
than unity (taking the relevant reference volumes to be string scale
in size). In this regime, one need only keep the leading terms in an
expansion in the inverse of the real parts of these moduli. This is
precisely the gaugino condensation limit. As discussed above, the
component action approach can then successfully reproduce the
supersymmetric potential energy using the simple form for the
condensate given in~\eqref{condensate}.

The above result was derived assuming the gauge coupling depends on 
the dilaton $S$ only. In this case, the gauge kinetic function is given by
\begin{equation}
f=S\;.
\label{kf}
\end{equation}
It will be useful to rewrite superpotential~\eqref{wgaugino} as
\bea \label{wgaugino2} W_{\textnormal{gaugino}} = A e^{-\epsilon f}
\;. \eea We now want to extend this result to the cases where there
are threshold corrections, including those due to supersymmetric
five-branes. In these circumstances, it is well-known
\cite{Derendinger:2000gy,Brandle:2001ts,mattias} that the gauge
kinetic functions on the $(0)$ and $(N+1)$ boundary walls generalize
to \bea \label{gkf02} f_{(0)} &=&S-\epsilon_0\left[
  \tau^{(N+1)}_{k}T^{k}+2\sum_{\hat{p}=1}^{N}Z_{(\hat{p})}\right],
\\ \label{gkfN12} f_{(N+1)} &=& S+
\epsilon_0\left[\tau_{k}^{(N+1)}T^{k}\right].\eea Here $T^k$ are the
$h^{1,1}$ K\"ahler moduli, the $\hat{p}$ index runs over all $N-1$
supersymmetric five-branes but excludes the anti-brane, and
$Z_{(\hat{p})}$ are the location moduli of these five-branes.  The complex fields $T^{k}$ and $Z_{(\hat{p})}$ are defined in~\eqref{superfields} of Appendix A. The
generalization of the gaugino condensate superpotential is now
straightforward. It is simply given by expression~\eqref{wgaugino2},
where $f$ takes the form~\eqref{gkf02} and~\eqref{gkfN12} on the $(0)$
and $(N+1)$ boundary walls respectively.

One can now find the gaugino condensate in the presence of threshold
effects by comparing expression~\eqref{componentresult} with the
complete potential energy derived from this modified
superpotential. As above, this turns out to be difficult in the
general case of arbitrary moduli expectation values. Once
again, the problem greatly simplifies in the gaugino condensate limit
of low energy and large moduli values. The threshold corrections do,
however, further complicate the polynomial prefactor.  Happily, in the
physically interesting region of moduli space further simplification
is possible. Note, using the definition of the superfields $Z_{(\hat{p})}$ in
equation \eqref{superfields} of Appendix A, that each of the two types of threshold 
corrections in equations \eqref{gkf02} and \eqref{gkfN12}
is suppressed relative to the leading dilaton term by factors of
$\epsilon_S$, as defined in~\eqref{BB},~\eqref{AA} and~\eqref{good}. Therefore,
since we require $\epsilon_S<<1$ for the validity of the
four-dimensional effective theory, we find that \bea
 \label{gkfN13}
 S &>& \epsilon_0\left[\tau_{k}^{(N+1)}T^{k}\right]\;,  \\
\label{gkf03}
S &>& \epsilon_0\left[2\sum_{\hat{p}=1}^{N}Z_{(\hat{p})} \right]\;.  \eea It follows
that, in the polynomial prefactor, it is a very good approximation to
simply ignore the threshold corrections, yielding the same prefactor
as discussed previously. Note, however, that even though, 
conditions~\eqref{gkfN13} and~\eqref{gkf03} strongly hold, the values
of the right-hand sides of these expressions are generally much larger
than unity. Hence, one should never drop the threshold and five-brane
corrections to $f$ in the exponential.

 Putting this all together, we conclude that in the gaugino
 condensate, $\epsilon_S<<1$ limit, the gaugino condensate is given by
 \bea \label{condensate2}{\bar{ X}}^{\Lambda}_y = i e^{{\kappa_4^2
     \cal K} ({\mathfrak z})} \frac{{A}}{\sqrt{2}} \kappa_4^2 \epsilon
 (S+\bar{S}) e^{-\epsilon f}, \eea where the gauge kinetic function
 $f$ is \bea \label{gkf04} f_{(0)} &=&S-\epsilon_0\left[
   \tau^{(N+1)}_{k}T^{k}+2\sum_{\hat{p}=1}^{N}Z^{(\hat{p})}\right],
 \\ \label{gkfN14} f_{(N+1)} &=& S+
 \epsilon_0\left[\tau_{k}^{(N+1)}T^{k}\right]\eea for a condensate on
 the $(0)$ and $(N+1)$ boundary walls respectively.

\subsection{Condensate scales II: anti-brane case}

Let us now turn to the case of heterotic M-theory in the presence of
anti-branes. At first glance, it is not obvious how to proceed. As we
have just seen, the arguments used in a discussion of gaugino
condensation are firmly rooted in the assumption that the theory is
supersymmetric. How, then, does one proceed when supersymmetry is
broken by anti-branes? The key observation we will make is that when
one takes the low energy, large modulus limit of a theory with
anti-branes, the system returns to a supersymmetric form. Hence, one
can continue to apply the usual arguments for gaugino condensation.

In the gaugino condensation limit, two types of terms survive in the Yang-Mills sector of the low
energy effective action. These are the kinetic terms for the gauge
fields and those for the gauginos. We presented the gauge field
kinetic terms, including the contribution of the anti-brane, in \cite{paper1}. These are of the same {\it form} as in the supersymmetric case.
However, the gauge kinetic functions for the $(0)$ and $(N+1)$ boundary walls are now given by
\bea \label{gkf05} f_{(0)} &=&S-\epsilon_0\left[
    \tau^{(N+1)}_{k}T^{k}+2\sum_{p=1}^{N}Z^{(p)} - \frac{2}{3}\delta_k
    (T^k + \bar{T}^k) \right. \\ && \;\;\;\; \quad\quad \quad \quad \quad
  \left. \nonumber + \delta_k (T^k-\bar{T}^k) \left(
      \left(\frac{Z_{(\bar{p})}+ \bar{Z}_{(\bar{p})}}{\bar{\tau}_k
          (T^k +\bar{T}^k)} \right)^2-2
      \frac{Z_{(\bar{p})}+\bar{Z}_{(\bar{p})}}{\bar{\tau}_k (T^k
          +\bar{T}^k)}
      \right) \right] ,\\ \label{gkfN15}
  f_{(N+1)} &=& S+
  \epsilon_0\left[\tau_{k}^{(N+1)}T^{k}-\frac{1}{3}\delta_k (T^k+
    \bar{T}^k) - \delta_k (T^k-\bar{T}^k) \left(\frac{Z_{(\bar{p})}+
        \bar{Z}_{(\bar{p})}}{\bar{\tau}_k (T^k +\bar{T}^k)} \right)^2
  \right] \eea
respectively. Similarly, we find that the gaugino kinetic terms are of the same form as in the supersymmetric case, but with the gauge kinetic functions replaced by expressions~\eqref{gkf05} and~\eqref{gkfN15}. 

The reason the gaugino kinetic terms retain their supersymmetric form is the
following. In the five-dimensional theory, there is only one set of kinetic terms
for the gauginos. These are localized on the appropriate orbifold
fixed plane. Upon dimensional reduction, there are three possible
sources of four-dimensional kinetic terms for these fields. The first
is the direct dimensional reduction of the five-dimensional kinetic
terms. The second arises when the contribution to the warping of the
bulk fields, which is proportional to the gaugino kinetic term, is
substituted into the tension terms of the various extended
objects. Finally, a contribution arises when the warping terms due to
the tension of the extended objects and the gaugino fluctuations on
the fixed planes are substituted into the bulk action. A simple
argument reveals that two of these contributions always cancel.

Consider the simple system of just the bulk action and
the tension terms on the extended objects. This action has a reduction
ansatz given by the warped anti-brane background presented
in \cite{paper1} and in expressions  (\ref{a})-(\ref{bk}) above. Now add, as a
perturbation, the gaugino terms in the action and the correction they
give rise to in the reduction ansatz. Substituting the reduction
ansatz into the action and integrating, we obtain the same four-dimensional 
action as before plus the new four-dimensional gaugino
kinetic terms. The term which arises from substituting the new warping piece 
into the zeroth order action clearly vanishes. This is because, by definition,
the zeroth order background extremizes the action in
the absence of gauginos. One is left with the direct
reduction of the five-dimensional gaugino kinetic term as the four-dimensional 
kinetic term for these fields. This reproduces exactly the supersymmetric  gaugino kinetic terms
written, however,  in terms of the gauge kinetic function~\eqref{gkf05} and~\eqref{gkfN15}. 

We conclude that the action for the gauge theories on the boundaries
is, in the gaugino condensation limit, of exactly the same {\it form}
as in globally supersymmetric Yang-Mills theory. The gauge kinetic
functions which appears in all of the kinetic terms are, however, given
by the non-holomorphic combination of moduli derived in \cite{paper1}
and presented in~\eqref{gkf05} and~\eqref{gkfN15}. Because of this
non-holomorphicity, the Yang-Mills sector of the theory is not, in
general, supersymmetric. However, in any situation where
the moduli are treated as constant, that is, independent of
space-time, the Yang-Mills sector is supersymmetric. The
non-holomorphy of the gauge kinetic function is not manifest if the
moduli, and, hence, the gauge coupling, are simply regarded as
numbers. Therefore, in this region of moduli space one can apply
exactly the same analysis of the gauge condensate as in the
supersymmetric case discussed in the previous subsection. That is, one simply
needs to replace $f$ by the correct gauge kinetic function for the
case at hand. We conclude, therefore, that in the gaugino condensate,
$\epsilon_S<<1$ limit, the condensate in the presence of anti-branes
is given by \bea \label{condensate22}{\bar{X}}^{\Lambda}_y = i
e^{{\kappa_4^2 \cal K} ({\mathfrak z})} \frac{{A}}{\sqrt{2}}
\kappa_4^2 \epsilon (S+\bar{S}) e^{-\epsilon f}, \eea where the gauge
kinetic function $f$ is \bea \label{gkf06} f_{(0)}
&=&S-\epsilon_0\left[ \tau^{(N+1)}_{k}T^{k}+2\sum_{p=1}^{N}Z^{(p)} -
  \frac{2}{3}\delta_k (T^k + \bar{T}^k) \right. \\ && \;\;\;\;
\quad\quad \quad \quad \quad \left. \nonumber + \delta_k
  (T^k-\bar{T}^k) \left( \left(\frac{Z_{(\bar{p})}+
        \bar{Z}_{(\bar{p})}}{\bar{\tau}_k (T^k +\bar{T}^k)}
    \right)^2-2 \frac{Z_{(\bar{p})}+\bar{Z}_{(\bar{p})}}{\bar{\tau}_k
      (T^k +\bar{T}^k)} \right) \right] ,\\ \label{gkfN16} f_{(N+1)}
&=& S+ \epsilon_0\left[\tau_{k}^{(N+1)}T^{k}-\frac{1}{3}\delta_k (T^k+
  \bar{T}^k) - \delta_k (T^k-\bar{T}^k) \left(\frac{Z_{(\bar{p})}+
      \bar{Z}_{(\bar{p})}}{\bar{\tau}_k (T^k +\bar{T}^k)} \right)^2
\right] \eea for a condensate on the $(0)$ and $(N+1)$ boundary walls
respectively.

This condensate can now be substituted into \eqref{componentresult} to
give the combined potential due to flux and
gaugino condensation for heterotic M-theory in the presence of
anti-branes. The result is
\bea V_{\textnormal{c+c/f+f}} = \frac{1}{2} e^{\kappa_4^2 K_{\textnormal{\rm mod}}}
  \left( \kappa_4^2 | A \epsilon (S+\bar{S}) e^{-\epsilon f} |^2 
 + \bar{A} \kappa_4^2 \epsilon (S+\bar{S}) e^{-\epsilon
      \bar{f}} W_{\textnormal{flux}} 
      + \kappa_4^2
    \bar{W}_{\textnormal{flux}} A \epsilon (S+\bar{S}) e^{-\epsilon f} \right) 
 \eea   

\bea  \nonumber
+e^{\kappa_4^2
    K_{\textnormal{\rm mod}}} \left( K_{\textnormal{\rm mod}}^{i
      \bar{j}} D_i W_{\textnormal{flux}} D_{\bar{j}}
    \bar{W}_{\textnormal{flux}} - 3 \kappa_4^2
    |W_{\textnormal{flux}}|^2 \right)  \eea 
In this expression
$W_{\textnormal{flux}}$ is given in \eqref{Wflux3}, $f$ is presented in
(\ref{gkf06}),(\ref{gkfN16}) and $K_{\textnormal{mod}}$ is defined in~\eqref{modk} of
Appendix A.

\section{Conclusions}
\label{conclusion}

In this paper, we have included the effects of flux and gaugino
condensation in the four-dimensional effective description of
heterotic M-theory including anti-branes \cite{paper1}. While the
parts of the resulting potential which are due purely to flux are
unchanged from the supersymmetric result, those which are caused by
gaugino condensation are modified in important ways.

It is not even obvious, a priori, that gaugino condensation would
occur in such a non-supersymmetric setting. However, because in a
certain limit the system still looks like globally supersymmetric
gauge theory, we have argued that indeed it does. We have also argued
that we can calculate an approximation to the potential for the moduli
which it gives rise to. It should be noted that, in addition to the
points explicitly mentioned in the proceeding sections, the threshold
corrections which anti-branes give rise to in the gauge kinetic
functions can completely change which extended
objects in the higher dimensional theory are strongly coupled - and so,
which exhibit gaugino condensation.

Let us summarize the results derived in Sections 4 and 5. We have shown that the moduli potential energy that arises from 1) perturbative
effects, 2) gaugino condensation and 3) flux in heterotic M-theory vacua with both M5-branes and anti M5-branes is, to our order of approximation,  given by \bea \label{fullpot}
V &=& {\cal V}_1 +  {\cal V}_2 +  V_{c+c/f+f} \\
{\cal{V}}_{1} &=& \frac{\epsilon_0 \, \kappa_4^{-2}}{(\pi \rho)^2}
(T^k+\bar{T}^k) \delta_k e^{\kappa_4^2 (K_T+K_D)} \;.  \\
{\cal{V}}_{2} &=& \frac{\epsilon_0^2 \, \kappa_4^{-4}}{(\pi \rho)^2 }
e^{\kappa_4^2 (K_T +2 K_D)} K_T^{\bar{k} l} \delta_l \left[
  \sum_{p=0}^{\bar{p}-1} \tau^{(p)}_k \frac{Z_{(\bar{p})} +
    \bar{Z}_{(\bar{p})}}{\bar{\tau}_m (T^m +\bar{T}^m)} -
  \sum_{p=\bar{p}+1}^{N+1} \tau^{(p)}_k \frac{Z_{(\bar{p})} +
    \bar{Z}_{(\bar{p})}}{\bar{\tau}_m (T^m +\bar{T}^m)} \right. \\
&&\left. \nonumber - \sum_{p=0}^{\bar{p}-1} \tau^{(p)}_k \frac{Z_{(p)}
    + \bar{Z}_{(p)}}{\tau_m^{(p)} (T^m +\bar{T}^m)} +
  \sum_{p=\bar{p}+1}^{N+1} \tau^{(p)}_k \frac{Z_{(p)} +
    \bar{Z}_{(p)}}{\tau_m^{(p)} (T^m +\bar{T}^m)} \right. \\ &&
\left. \nonumber + \sum_{p=0}^{N+1} \tau^{(p)}_k \left( 1-
    \frac{Z_{(p)}+\bar{Z}_{(p)}}{\tau^{(p)}_m (T^m+\bar{T}^m)} \right)
  \frac{Z_{(p)}+\bar{Z}_{(p)}}{\tau^{(p)}_n (T^n+\bar{T}^n)}
  -\frac{2}{3} \delta_k \right] \\
V_{c+c/f+f} &=& \left[ e^{ \kappa_4^2 K_{\textnormal{mod}}}
  \left( \kappa_4^2 | A \epsilon (S+\bar{S}) e^{-\epsilon f} |^2 + \bar{A} \epsilon
    (S+\bar{S}) e^{-\epsilon \bar{f}} \kappa_4^2 W_{\textnormal{flux}} +
    \kappa_4^2 \bar{W}_{\textnormal{flux}} A \epsilon (S+\bar{S}) e^{-\epsilon f}
  \right) \right. \\ \nonumber && \left.  + e^{\kappa_4^2
    K_{\textnormal{mod}}} \left( K_{\textnormal{mod}}^{i
      \bar{j}} D_i W_{\textnormal{flux}} D_{\bar{j}}
    \bar{W}_{\textnormal{flux}} - 3 \kappa_4^2
    |W_{\textnormal{flux}}|^2 \right) \right] \eea Here the
superpotential for the flux is given by the expression \bea
W_{\textnormal{flux}} = \frac{\sqrt{2}}{\kappa_4^2} \int_X \Omega \wedge G=
\frac{\sqrt{2}}{\kappa_4^2} \left( {\cal X}^A
  {\cal G}_A - \tilde{{\cal X}}_B {\cal Z}^B \right) 
 \; ,\eea the
appropriate gauge kinetic function, $f$, should be chosen from
\eqref{gkf05} or \eqref{gkfN15} and the K\"ahler potential
$K_{\textnormal{mod}}$ is defined in~\eqref{modk} of Appendix A.

The contribution of non-perturbative membrane instanton effects will be added to this potential
in future work, as will an analysis of the vacua of the
resulting system \cite{paper3}.

\vskip 0.5cm
\noindent
{\bf Acknowledgments\\}
A.~L.~is supported by the EC 6th Framework Programme
MRTN-CT-2004-503369. B.~A.~O.~is supported in part by the DOE under
contract No. DE-AC02-76-ER-03071 and by the NSF Focused Research Grant
DMS0139799.  J.~G.~is supported by CNRS.  J.~G.~and A.~L.~would like
to thank the Department of Physics, University of Pennsylvania where
part of this work was being carried out for generous hospitality. 

\section*{Appendix A: Field Definitions and K\"ahler Potentials for
  Heterotic Vacua with Anti-Branes}

In this Appendix, we briefly state some results from \cite{paper1}
which are required in the main text. Despite the explicit
supersymmetry breaking introduced by the anti M5-brane in our vacuum, the
kinetic terms in the four-dimensional theory can still be be expressed, as in the supersymmetric case, 
in terms of a K\"ahler potential and complex structure. Let us define the complex scalar fields
\bea \label{superfields} S
&=&e^\phi+i\sigma +\epsilon_0\sum_{p=1}^{N} \tau_{k}^{(p)}z_{(p)}^2T^{k} \\
\nonumber T^{k} &=& e^\beta b_{0}^{k}+2i\chi^{k} \\ \nonumber 
Z_{(p)} 
&=&z_{(p)}
\tau^{(p)}_{k}T^{k}-2i\tau_{k}^{(p)}n^{k}_{(p)}\nu_{(p)}, \, \eea 
where $V_{0}=e^{\phi}$,  $b_{0}=e^{\beta}$,  $b_{0}^{k}$ and
$z_{(p)}$ are specified in Section 2 and  $\sigma$, $\chi^{k}$ and $\nu_{(p)}$ are their real scalar superpartners given in~\cite{paper1}. Note that in the case with no anti branes, each of these complex scalars would be the lowest component of an ${\cal{N}}=1$ chiral superfield. This is no longer the case when the vacuum contains an anti M5-brane. Be that as it may, we find that the kinetic terms of the bosonic sector of heterotic M-theory in the presence of anti-branes is still given, in terms of these fields, by the usual ${\cal{N}}=1$ supersymmetric
formula with the appropriate K\"ahler potentials.  Up to order $\kappa_{11}^{2/3}$, these K\"ahler potentials are found to be
\begin{equation}\label{scalKaehlerPot}
   \k_{4}^{2}\, K_{\rm scalar}=\kappa_4^2 K_{\rm  mod} + \kappa_4^2 K_{\rm matter}\; ,
\end{equation}
where
\begin{equation}
K_{\rm mod}=K_{D}+K_{T}+\cK
\label{modk}
\end{equation}
and
\begin{eqnarray}\label{K}
  \kappa_4^2 K_{D}&=&-{\rm ln}\left[S+\bar{S}-\epsilon_0\sum_{p=1}^{N}
    \frac{(Z_{(p)}+\bar{Z}_{(p)})^{2}}{\tau_{k}^{(p)}(T^{k}+\bar{T}^{k})}
  \right]\,,\\
  \kappa_4^2 K_{T}&=&-{\rm ln}\left[\frac{1}{48}d_{klm}(T^{k}+\bar{T}^{k})(T^{l}+\bar{T}^{l})
    (T^{m}+\bar{T}^{m})\right] \\
  \kappa_4^2 
\cK({\mathfrak z})&=&- \mbox{ln} \left[2 i ({\mathcal G}-\bar{{\mathcal G}})
    - i ( \mathfrak{z}^{\underline{a}}-\bar{\mathfrak{z}}^{\bar{\underline{a}}})\left(\frac{\pt
        \mathcal{G}}{\pt\mathfrak{z}^{\underline{a}}}+\frac{\pt
        \bar{\mathcal{G}}}{\pt\bar{\mathfrak{z}}^{\bar{\underline{a}}}}\right) \right]\\
  K_{\rm matter}&=&e^{\kappa_4^2 K_{T}/3}\sum_{p=0,N+1} G_{(p)MN}
  C^{Mx}_{(p)}\bar{C}_{(p)x}^{N} \; .
\end{eqnarray}
The symbol ${\cal{G}}$ in $\cK({\mathfrak z})$ is the ${\cal{N}}=2$ prepotential of the $h^{2,1}$ sector. It is defined in terms of the periods ${\cal{G}}_{\underline{A}}$ 
in~\eqref{amore} by ${\cal{G}}_{\underline{A}}=\frac{\partial}{\partial{\cal{Z}}^{\underline{A}}}{\cal{G}}$. The significance of the $\kappa_{11}^{\frac{2}{3}}$ expansion relative
to that in $\epsilon_S$ is described in \cite{paper1}.

\section*{Appendix B: Reproducing the Flux Potential from the Gukov-Vafa-Witten
  Superpotential}
  
 Here we present a proof that, in heterotic M-theory vacua with $N-1$ M5-branes and an anti M5-brane, the Gukov-Vafa-Witten superpotential
\begin{equation}
  W_{\rm flux}=\frac{\sqrt{2}}{ \k_{4}^{2}}\;(\cX^{\underline{A}}\cG_{\underline{A}}-\tilde{\cX}_{\underline{A}}\cZ^{\underline{A}}),
\end{equation}
along with K\"ahler potential $K_{\rm mod}$ given in~\eqref{modk},
reproduces the four-dimensional scalar potential energy~\eqref{fullpotential2} when they are inserted into the {\it supersymmetric} expression for the potential. The manipulations presented below were first presented in~\cite{mattias} in the context of vacua without anti branes. The scalar potential energy generated by $W_{\rm flux}$ and $K_{\rm mod}$ is
\begin{equation}
   V_{\rm flux}=e^{\k_{4}^{2}K_{\rm mod}}\left(K^{i\bar{j}}_{\textnormal{mod}} D_{i}W_{\rm flux}\overline{D_{j}W_{\rm flux}} -3\k_{4}^{2}|W_{\rm flux}|^{2}\right)\,,
\end{equation} where all of the complex fields 
\eqref{superfields} are collectively denoted 
by $Y^{i}$ and the K\"ahler covariant 
derivative is $D_{i}W=\pt_{i}W+\k^{2}_{4}\frac{\pt K_{\textnormal{scalar}}}{\pt 
  Y^{i}}W$. Using the form of the K\"ahler potential and the 
fact that the superpotential depends only on the complex 
structure moduli $\mathfrak{z}^{a}$, it can be shown that
\begin{eqnarray}
  V_{\rm flux}&=&e^{\kappa_4^2( K_T + K_D)}e^{\kappa_4^2 \cK}\left[\cK^{\underline{a}\bar{\underline{b}}}D_{\underline{a}}W_{\rm flux}\overline{D_{\underline{b}}W_{\rm flux}}+\left(\kappa_4^4 \tilde{K}^{u\bar{v}}
      \pt_{u} \tilde{K}\overline{\pt_{v} \tilde{K}}-3 \kappa_4^2 \right)|W_{\rm flux}|^{2}\right] \\
  &=&\frac{1}{2\,b_{0}^{3}V_{0}}e^{\kappa_4^2 \cK}\left[\cK^{\underline{a}\bar{\underline{b}}}D_{\underline{a}}W_{\rm flux}
    \overline{D_{\underline{b}}W_{\rm flux}}+\k_{4}^{2}|W_{\rm flux}|^{2}\right]\;.
\end{eqnarray}
Here, we have defined $\tilde{K}\equiv K_{D}+K_{T}$, the subscripts $u,v$ run over all moduli indices except for $\underline{a},\bar{\underline{a}}$, and we have used the fact that
$\tilde{K}^{u\bar{v}}\pt_{u}\tilde{K}\overline{\pt_{v} \tilde{K}}=4
\kappa_4^{-2}$. As in \cite{paper1}, $\cK$ is the K\"ahler potential of
the complex structure moduli. We now use
$\pt_{\underline{a}}=\frac{\pt}{\pt\mathfrak{z}^{\underline{a}}}
=\frac{\pt\cZ^{\underline{A}}}{\pt\mathfrak{z}^{\underline{a}}}\frac{\pt}{\pt
  \cZ^{\underline{A}}}=\pt_{\underline{a}}\cZ^{\underline{A}}\pt_{\underline{A}}$ and $\cK_{\underline{A}}=\pt_{\underline{A}}\cK$ to write
\begin{eqnarray}
  \cK^{\underline{a}\underline{\bar{b}}}D_{\underline{a}}W_{\rm flux}\overline{D_{\underline{b}}W_{\rm flux}}&=&(\cK^{\underline{a}\underline{\bar{b}}}\pt_{\underline{a}}
  \cZ^{\underline{A}}\pt_{\underline{\bar{b}}}\bar{\cZ}^{\underline{B}})
  (\pt_{\underline{A}}W_{\rm flux}+ \kappa_4^2 \cK_{\underline{A}}W_{\rm flux})(\overline{\pt_{\underline{B}}W_{\rm flux}+ \kappa_4^2 \cK_{\underline{B}}W_{\rm flux}})\\
  &=& \frac{2}{\k_{4}^{2}}\;(\cK^{\underline{a}\underline{\bar{b}}}\pt_{\underline{a}}\cZ^{\underline{A}}\pt_{\underline{\bar{b}}}\bar{\cZ}^{\underline{B}})
  \left[(\cX^{\underline{C}}\cG_{\underline{C}\underline{D}}-\tilde{\cX}_{\underline{D}})(\d^{\underline{D}}_{\underline{A}}+\cZ^{\underline{D}} 
    \kappa_4^2 {\cal K}_{\underline{A}})\right]\\
  &&\qquad\qquad\qquad\;\; \times
  \left[(\cX^{\underline{E}}\bar{\cG}_{\underline{E}\underline{F}}-\tilde{\cX}_{\underline{F}})(\d^{\underline{F}}_{\underline{B}}+ \kappa_4^2 
    \overline{\cZ^{\underline{F}}
      {\cal K}_{\underline{B}}})\right]\nn\\
  &=& \frac{2}{\k_{4}^{2}}\;(\cX^{\underline{C}},\tilde{\cX}_{\underline{C}})\left[\begin{array}{cc}
      \cG_{\underline{C}\underline{D}}U^{\underline{D}\underline{F}}\bar{\cG}_{\underline{F}\underline{E}}\quad & -\cG_{\underline{C}\underline{D}}U^{\underline{D}\underline{E}} \\
      -U^{\underline{C}\underline{F}}\bar{\cG}_{\underline{F}\underline{E}} & U^{\underline{C}\underline{E}} \end{array}\right]{\cX^{\underline{E}}
    \choose \tilde{\cX}_{\underline{E}}}
\end{eqnarray}Here, $U^{\underline{D}\underline{F}}$ is given by the expression
\begin{equation}
  U^{\underline{D}\underline{F}}=(\cK^{\underline{a}\underline{\bar{b}}}\pt_{\underline{a}}\cZ^{\underline{A}}\pt_{\underline{\bar{b}}}\bar{\cZ}^{\underline{B}})
  (\d^{\underline{D}}_{\underline{A}}+\kappa_4^2 \cZ^{\underline{D}}{\cal K}_{\underline{A}})
  (\d^{\underline{F}}_{\underline{B}}+\kappa_4^2 \overline{\cZ^{\underline{F}}{\cal K}_{\underline{B}}})\;.
\end{equation}
We now define
\begin{equation}\label{M-matrix}
  M_{\underline{A}\underline{B}}=\bar{\cG}_{\underline{A}\underline{B}}+T_{\underline{A}\underline{B}}\,,\qquad
  T_{\underline{A}\underline{B}}=2i\frac{\mbox{Im}\cG_{\underline{A}\underline{C}}{\cal Z}^{\underline{C}}\;\mbox{Im}\cG_{\underline{B}\underline{D}}{\cal Z}^{\underline{D}}}
  {{\cal Z}^{\underline{E}}\mbox{Im}\cG_{\underline{E}\underline{F}}{\cal Z}^{\underline{F}}}\; .
\end{equation}
One  can then use the relations \bea U^{\underline{A}\underline{B}} M_{\underline{B}\underline{C}} = U^{\underline{A}\underline{B}} \bar{{\cal
    G}}_{\underline{B}\underline{C}} \;, \;\;\; {\cal G}_{\underline{A}\underline{B}} U^{\underline{B}\underline{C}} = \bar{M}_{\underline{A}\underline{B}} U^{\underline{B}\underline{C}} \;,
\eea together with the explicit form for $U^{\underline{D}\underline{E}}$, \bea U^{\underline{A}\underline{B}} =
-\frac{1}{2} e^{-\kappa_4^2 {\cal K}} (\textnormal{Im} M)^{-1 \underline{A}\underline{B}} - \bar{{\cal
    Z}}^{\underline{A}} {\cal Z}^{\underline{B}}\;, \eea to write,
\begin{eqnarray}
 &&\left[\begin{array}{cc}
   \cG_{\underline{C}\underline{D}}U^{\underline{D}\underline{F}}\bar{\cG}_{\underline{F}\underline{E}}\quad & -\cG_{\underline{C}\underline{D}}U^{\underline{D}\underline{E}} \\
   -U^{\underline{C}\underline{F}}\bar{\cG}_{\underline{F}\underline{E}} & U^{\underline{C}\underline{E}} \end{array}\right] =\left[\begin{array}{cc}
   \bar{M}_{\underline{C}\underline{D}}U^{\underline{D}\underline{F}}M_{\underline{F}\underline{E}}\quad & -\bar{M}_{\underline{C}\underline{D}}U^{\underline{D}\underline{E}} \\
   -U^{\underline{C}\underline{F}}M_{\underline{F}
   \underline{E}} & U^{\underline{C}\underline{E}} \end{array}\right] \\
   &=&-\frac{e^{-\kappa_4^2 \cK}}{2}\left[\begin{array}{cc}
   \bar{M}_{\underline{C}\underline{D}}({\rm Im}M^{-1})^{\underline{D}\underline{F}}M_{\underline{F}\underline{E}}\quad & -\bar{M}_{\underline{C}\underline{D}}({\rm Im}M^{-1})^{\underline{D}\underline{E}} \\-({\rm Im}M^{-1})^{\underline{C}\underline{F}}M_{\underline{F}\underline{E}} & ({\rm Im}M^{-1})^{\underline{C}\underline{E}} \end{array}\right]- \left[\begin{array}{cc}
   \bar{\cG}_{\underline{C}}\cG_{\underline{E}}\quad & -\bar{\cG}_{\underline{C}}\cZ^{\underline{E}} \\
   -\bar{\cZ}^{\underline{C}}\cG_{\underline{E}} & \bar{\cZ}^{\underline{C}}\cZ^{\underline{E}}    \end{array}\right]\nn \;.
\end{eqnarray} 
In the last matrix, the relation $(M_{\underline{A}\underline{B}} - \bar{{\cal
    G}}_{\underline{A}\underline{B}}){\cal Z}^{\underline{B}} = 2 i \textnormal{Im} {\cal G}_{\underline{A}\underline{B}} {\cal Z}^{\underline{B}}
$ has also been employed. Finally, using the identity
\begin{equation}
  |W_{\rm flux}|^{2}=\frac{2}{\k_{4}^{4}}\;(\cX^{\underline{C}},\tilde{\cX}_{\underline{C}})\left[\begin{array}{cc}
   \bar{\cG}_{\underline{C}}\cG_{\underline{E}}\quad & -\bar{\cG}_{\underline{C}}\cZ^{\underline{E}} \\
   -\bar{\cZ}^{\underline{C}}\cG_{\underline{E}} & \bar{\cZ}^{\underline{C}}\cZ^{\underline{E}} \end{array}\right] {\cX^{\underline{E}}
   \choose \tilde{\cX}_{\underline{E}}}\;,
\end{equation}
the scalar potential becomes
\begin{eqnarray}
  V_{\rm flux}&=&-\frac{1}{2\k_{4}^{2}\,b_{0}^{3}V_{0}}(\cX^{\underline{C}},\tilde{\cX}_{\underline{C}})\left[\begin{array}{cc}
      \bar{M}_{\underline{C}\underline{D}}({\rm Im}M^{-1})^{\underline{D}\underline{F}}M_{\underline{F}\underline{E}}\quad & -\bar{M}_{\underline{C}\underline{D}}({\rm Im}M^{-1})^{\underline{D}\underline{E}} \\ -({\rm Im}M^{-1})^{\underline{C}\underline{F}}M_{\underline{F}\underline{E}} & ({\rm Im}M^{-1})^{\underline{C}\underline{E}} \end{array}\right]{\cX^{\underline{E}}
    \choose \tilde{\cX}_{\underline{E}}}.\nn\\
\end{eqnarray}
This is easily rewritten as
\begin{equation}
V_{\rm flux}=-\frac{1}{2\k_{4}^{2}\,b_{0}^{3}V_{0}}(\tilde{\cX}_{\underline{A}}-\bar{M}_{\underline{A}\underline{B}}\cX^{\underline{B}})
 [\mbox{Im}M]^{-1\,\underline{A}\underline{C}}(\tilde{\cX}_{\underline{C}}-
 M_{\underline{C}\underline{D}}\cX^{\underline{D}}),
 \label{what}
 \end{equation}
which is exactly the four-dimensional flux potential~\eqref{fullpotential2} obtained by dimensional reduction. This completes the proof. We emphasize again that this proof was carried out in the presence of both M5-branes and anti M5-branes in the vacuum. It is, of course, valid in the purely supersymmetric case as well.

\vspace{0.1cm}

Finally, we take the opportunity to state some identities that we use
in the body of the paper.  These are\bea
\label{ident1}{\cal G}_{\underline{A}} = M_{\underline{A}\underline{B}} {\cal Z}^{\underline{B}} \\
\label{ident2} \textnormal{Im} M_{\underline{A}\underline{B}} {\cal Z}^{\underline{A}} \bar{{\cal Z}}^{\underline{B}} =
-\frac{1}{2} e^{- \kappa_4^2 {\cal K}} \eea


\end{document}